\documentclass[aps,prb,superscriptaddress,showpacs,twocolumn]{revtex4}
\usepackage{graphicx}
\usepackage{amssymb}
\usepackage{amsmath}

\newcommand{\bk}{{\bf k}}
\newcommand{\bq}{{\bf q}}
\newcommand{\bp}{{\bf p}}
\newcommand{\bx}{{\bf x}}

\renewcommand{\Im}{{\mathop{\rm{Im}}\nolimits}}
\newcommand{\Gi}{\rm{Gi}}

\begin{document}

\title{Effects of an electronic topological transition for anisotropic
   low-dimensional superconductors}
 
\author{G. G. N. Angilella}
\affiliation{Dipartimento di Fisica e Astronomia, Universit\`a di
   Catania,\\ and Istituto Nazionale per la Fisica della Materia,
   UdR di Catania,\\ Corso Italia, 57, I-95129 Catania, Italy}
\author{E. Piegari}
\affiliation{DMFCI, Universit\`a di
   Catania, Viale A. Doria, 6, I-95125 Catania, Italy}
\affiliation{Istituto Nazionale per la Fisica della Materia,
   UdR di Firenze,\\ Via G. Sansone, 1, I-50019 Sesto Fiorentino (FI),
   Italy}
\author{A. A. Varlamov}
\affiliation{Istituto Nazionale per la Fisica della Materia,
   UdR di Tor Vergata,\\ DSTFE, Universit\`a di Roma ``Tor Vergata'',\\
   Via di Tor Vergata, 110, I-00133 Roma, Italy}
    
\date{\today}

\begin{abstract}
We study the superconducting properties of a two-dimensional superconductor in
   the proximity to an electronic topological transition (ETT).
In contrast to the 3D case, we find that the superconducting gap at
   $T=0$, the critical temperature $T_c$, and the impurity scattering rate
   are characterized by a nonmonotonic behavior, with maxima occurring
   close to the ETT.
We derive analytical expressions for the value of such maxima both in
   the $s$-wave and in the $d$-wave case.
Such expressions are in good qualitative agreement with the
   phenomenological trend recently observed for $T_c^{\rm{max}}$ as a
   function of the hopping ratio $t^\prime /t$ across several cuprate
   compounds.
We further analyze the effect of an ETT on the Ginzburg-Landau stiffness
   $\eta$.
Instead of vanishing at the ETT, as could be expected, thus giving
   rise to an increase of the fluctuation effects, in the case
   of momentum-independent electron-electron interaction, we find
   $\eta\neq 0$, as a result of an integration over the whole Fermi
   surface.\\
\pacs{%
74.20.-z,
74.62.-c,
74.40.+k
}
\end{abstract} 

\maketitle

\section{Introduction}

The unconventional properties of the normal and superconducting
   states of several low-dimensional novel electronic materials is a
   source of continuous interest and research.
Such materials include the high-$T_c$ cuprate superconductors
   (HTSC) \cite{Anderson:97},
   as well as some organic superconductors based on doped
   BEDT-TTF layers, and the ruthenates.
In these materials, the interplay between their reduced dimensionality
   and the strength of the effective electron-electron interaction is
   believed to be the key for the elusive nature of their normal
   state, as well as for the anisotropic gap characterizing their
   superconducting state.

A feature common to almost all the material classes listed
   above is a quasi-2D dispersion relation, arising from their layered 
   structure and stabilized by the tendency to confined coherence
   within layers, due to strong correlations
   \cite{Clarke:97}.
Indeed, flat bands have been observed in nearly all hole- and
   electron-doped superconductors \cite{Shen:95}, in the $\kappa$
   phase of BEDT-TTF organic superconductors \cite{McKenzie:97}, as
   well as in the noncuprate layered superconductor Sr$_2$RuO$_4$
   (Ref.~\onlinecite{Yokoya:96}).
Clear evidence for a 2D Fermi surface changing topology as a function
   of doping has been recently provided by ARPES measurements in LSCO
   \cite{Ino:01}.
In particular, the role of the proximity to an electronic
   topological transition in establishing the unconventional
   properties especially of the cuprates has been very early
   emphasized (see Ref.~\onlinecite{Markiewicz:97} for a
   review).
Therefore, in the following we will be mainly concerned with the case
   of the high-$T_c$ cuprates~\cite{note:RBCO}.

The term electronic topological transition (ETT) has been proposed in
   order to describe the phenomena related to a change of the
   connectivity number of the components of the Fermi surface (FS)
   \cite{Varlamov:89}. 
Such a transition can be driven by several causes such as isotropic
   pressure, anisotropic deformation, and the introduction of isovalent
   impurities. 
All these influences can be parametrized by their effect on the
   chemical potential $\mu$ passing through the critical value
   $\varepsilon_c$, corresponding to the transition point. 
Indeed, such a critical point can be well defined only in a pure
   metal at  $T=0$ where a true phase transition
   of order $2\frac{1}{2}$ occurs~\cite{Lifshitz:60},
   according to Ehrenfest classification.
Typical manifestations of an ETT consist in cusp-like anomalies of
   physical quantities such as the specific heat \cite{Dorbolo:98},
   the density of states (DOS), and the conductivity, as well as in
   the appearance of 
   asymmetric singularities of the thermal expansion coefficient and
   thermoelectric power in the dependence of all these quantities on
   $z=\mu -\varepsilon_c$. 
A non-zero temperature or the presence of electron scattering results in
   the smearing of these anomalies and, strictly speaking, in washing out
   the notion of a $2\frac{1}{2}$-order phase transition itself.
Moreover, the occurrence of an ETT can be masked by an intervening
   structural transition, as could be induced by external pressure.
The effects of an ETT on the properties of metals and alloys have been
   thoroughly investigated as well
   \cite{Varlamov:89,Blanter:94,Bruno:94}.

In lower dimensional metallic systems, an ETT is characterized by yet
   stronger anomalies.
In particular, the DOS of a 2D metal increases
   logarithmically near an ETT, instead of displaying a square-root
   cusp, as in the 3D case \cite{note:fluctuations}.
Therefore, it has been suggested that an ETT may
   be a clue for the understanding of the anomalous superconducting
   state of the high-$T_c$ cuprates \cite{Markiewicz:97}.
In particular, it is well known that the presence of an ETT in the
   spectrum of a 2D superconductor induces a nonmonotonic dependence
   of the critical temperature on doping or applied pressure
   \cite{Tsuei:90,Markiewicz:97}, in qualitative agreement with the
   available experimental results \cite{Zhang:93,Wijngaarden:99}.
This has to be contrasted with the 3D case, where an ETT only gives
   rise to a step-like behavior in the $z$-dependence of $T_c$
   \cite{Makarov:65}.

Moreover, it has been proposed that the proximity to an ETT may be the 
   origin of the unconventional normal state of the HTSC.
In particular, a marginal Fermi liquid
   \cite{Pattnaik:92,Newns:92,Newns:92a,Gopalan:92} or a non Fermi liquid
   \cite{Dzyaloshinskii:96} behavior can be naturally derived for a 2D 
   electron system near a Van~Hove singularity.
More generally, it has been argued that the anomalous
   finite-temperature phenomenology of the cuprates stems from the
   competition of several broken-symmetry states intervening near one
   and the same quantum critical point (QCP)
   \cite{Vojta:00,Chakravarty:01}.
Recently, Onufrieva \emph{et al.} \cite{Onufrieva:99a,Onufrieva:99b}
   showed that an ETT occurring in a 2D square lattice with hopping
   beyond nearest neighbors is a QCP, with
   two aspects of criticality: the first is related to the singular
   behavior of the thermodynamic properties (Van~Hove singularity),
   while the second is related to the existence of the critical line
   $T=0$, $z>0$ of static Kohn singularities
   \cite{Onufrieva:99a,Onufrieva:99b}.
The proximity to an ETT may be characterized by spin density wave (SDW),
   charge density wave (CDW), and $d$-wave superconducting (dSC)
   instabilities, depending on the appropriate interaction channels
   included in the analysis.
Onufrieva \emph{et al.} argued that SDW fluctuations dominate in the
   case of the high-$T_c$ cuprates \cite{Onufrieva:00}.
On the other hand, recent studies focussed on the competition between AFM
   and AFM-mediated $d$-wave pairing via a diagrammatic approach
   \cite{Onufrieva:96}, among dSC,
   AFM, and $\pi$-triplet pairing at a mean field level in the
   presence of backward scattering \cite{Murakami:98}, among dSC, AFM, 
   and CDW within the renormalization group (RG) approach
   \cite{Alvarez:98a,Alvarez:98,Gonzalez:00}, among dSC, AFM, and FM
   within the RG and the parquet approaches \cite{Irkhin:01}, or
   between dSC and an excitonic ordered state \cite{Kiselev:00}.
More recently, it has been shown \cite{Honerkamp:01} that elastic
   umklapp scattering near a Van~Hove singularity may give rise to an
   RVB-like, insulating spin liquid state \cite{Furukawa:98}, which
   exhibits both $d$-wave superconducting and AFM correlations,
   without being characterized by true symmetry breaking, as is
   typical of a quantum ordered state.
The competition between superconductivity and various kinds of density 
   waves in several low-dimensional electron systems in the presence of a 
   Van Hove singularity has been reviewed both from the experimental
   and the theoretical point of view in Ref.~\onlinecite{Gabovich:01}.

In this paper, we will concentrate on a single superconducting
   instability (towards either an $s$-wave or a $d$-wave superconducting
   state), in the weak coupling limit, thus neglecting altogether any
   other competing ordered phase, for a 2D electron system near an
   ETT.
This is of course justified only if all other instabilities are
   characterized by weaker couplings, which may not be the case for
   the cuprates.
However, such an approximation will enable us to derive an analytical
   expression for the maximum gap $\Delta_0$ near the ETT as a
   function of the band details.
Such results are in good qualitative agreement with recent studies of
   $T_c$ in the cuprates correlated with material dependent
   properties, such as the ratio of next-nearest to nearest neighbor
   hopping \cite{Pavarini:01}.

We will also study the effect of an ETT on the Ginzburg-Landau
   stiffness $\eta\propto\Gi^{-1}$, where $\Gi$ is the
   Ginzburg-Levanyuk parameter, which characterizes the manifestation range of
   fluctuations near $T_c$ (Ref.~\onlinecite{Varlamov:99}). 
In the case of an isotropic FS, $\eta$ is proportional to the square
   of the Fermi velocity.
In the vicinity of an ETT, due to the presence of `slow' electrons
   near the saddle point in the electronic spectrum, one may
   expect an increase of fluctuations.
However, we will show that, in the case of a momentum-independent
   electron-electron interaction, all electronic states on the FS
   participate in establishing the superconducting correlations.
Such correlations give rise to a superconducting stiffness, whose
   value is of the same order of magnitude of the result obtained for
   an isotropic electronic spectrum \cite{Perali:00}, to the lowest
   order in $z/E_{\mathrm{F}}$.

The paper is organized as follows.
In Sec.~\ref{sec:model}, we introduce a dispersion relation beyond
   nearest neighbors for an electron system in a 2D square lattice, as 
   is typical for the cuprates, and discuss its corresponding singular 
   DOS.
In Sec.~\ref{sec:super}, we study the superconducting gap $\Delta_0$ at
   $T=0$ as a function of the critical parameter $z$, in the case of
   $s$- and $d$-wave pairing.
In Sec.~\ref{sec:impurities}, we discuss the effect of impurities on
   the normal state DOS and show that the proximity to an ETT gives
   rise to a nonmonotonic $z$-dependence of the renormalized quasiparticle
   inverse lifetime $\tau^{-1}$.
In Sec.~\ref{sec:GL}, we calculate the Cooper pair propagator near an
   ETT, and discuss the effects of an ETT on the superconducting
   fluctuations.
We eventually summarize in Sec.~\ref{sec:conclusions}.

\section{The model}
\label{sec:model}

\begin{figure}[ht]
\centering
\includegraphics[height=0.9\columnwidth,angle=-90]{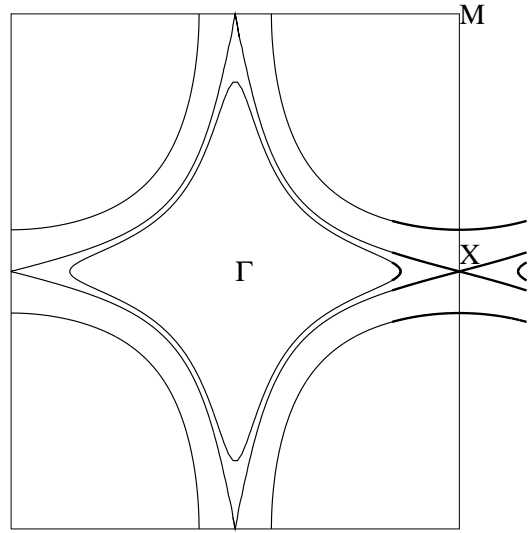} 
\caption{Fermi line $\xi_\bk = 0$, Eq.~(\protect\ref{eq:bdisp}), for a
   value of the hopping ratio $r=0.45$.
As the chemical potential varies from the bottom to the top of the
   band, the Fermi line changes topology, evolving from a closed
   contour around the $\Gamma$ point, to a contour, whose
   continuation in the higher order Brillouin zones closes around
   $M=(\pi,\pi)$.
The change of topology (ETT) occurs when the Fermi line touches the
   zone border, i.e. at $X=(\pi,0)$, and symmetry related points.
The thicker solid lines evidence the hyperbola-like shape of the Fermi
   line, Eq.~(\protect\ref{eq:hyperbola}), around $X$.
}
\label{fig:FS}
\end{figure}

Detailed band structure calculations within the local-density approximation
   (LDA)~\cite{Andersen:95}, as well as ARPES
   experiments~\cite{Randeria:99}, show that a realistic 
   tight-binding approximation for the band dispersion of most
   quasi-2D cuprates has to be expanded at least up to
   next-nearest neighbors hopping.
We then assume the following rigid band dispersion relation for a
   tetragonal lattice: 
\begin{equation}
\xi_\bk =\varepsilon_\bk -\mu =-2t(\cos k_x+\cos
   k_y)+4t^{\prime}\cos k_x\cos k_y-\mu ,   
\label{eq:bdisp}
\end{equation}
where $\mu $ denotes the chemical potential, and the components of
   the wavevector $\bk$ are measured in units of the inverse
   lattice spacing.
A non-zero value of the hopping ratio $r=t^\prime /t$, measuring the
   ratio of next-to-nearest vs nearest neighbors hopping, slightly
   modulates the actual shape of the Fermi line $\xi_\bk =0$,
   and in particular destroys perfect nesting at $\mu =0$ as well as
   electron-hole symmetry (Fig.~\ref{fig:FS}). 
In order to have a flat minimum in $\xi_\bk$ around the
   $\Gamma$ point, as is observed experimentally for the majority of
   the cuprates \cite{Dessau:93,Abrikosov:93,Gofron:94}, the condition
   $0<r< \frac{1}{2}$ must be fulfilled.
The role of an extended saddle point in stabilizing superconductivity
   against other possible low-energy instabilities has been
   established within the renormalization group (RG) approach in the
   weak-coupling limit \cite{Alvarez:98}.
Moreover, it has been shown that an increase of $r$ in the mentioned range
   correlates with an increase of the maximum
   $T_c$ across different classes of cuprate superconductors
   \cite{Pavarini:01}. 
However, changes of the \emph{shape} of the FS
   resulting from screening effects can also correlate with changes
   in the superconducting properties in an indirect way.
Indeed, a deformation of the FS
   also induces a change of the phase space effectively
   probed by the electron-electron interaction \cite{Hodges:71}. 
This is particularly relevant for several models proposed for the
   HTSC, characterized by effective interactions peaked at
   $X=(0,\pi )$, namely exactly where the
   Fermi line is most sensible to a change in the hopping range $r$.
Possible realizations of such a strongly anisotropic
   $\bk$-dependent potential include the interaction mediated by
   antiferromagnetic spin fluctuation \cite{Millis:90} or by
   quasicritical stripe fluctuations, due to the
   proximity to a QCP near optimal doping at $T=0$
   (Refs.~\onlinecite{Emery:93,Emery:95a,Castellani:95}), as well as
   electron-electron interactions enhanced by interlayer
   pair-tunneling (ILT) \cite{Chakravarty:93,Angilella:99}. 

As the chemical potential $\mu$ in Eq.~(\ref{eq:bdisp}) varies
   from the bottom, $\varepsilon_\perp =-4t (1-r)$, to the
   top of the band, $\varepsilon_\top =4t(1+r)$, the Fermi
   line $\xi_\bk =0$ evolves from an electron-like contour,
   closed around the $\Gamma$ point, to a hole-like contour, whose
   continuation into higher order Brillouin zones closes around the
   $M=(\pi ,\pi )$ point.
In doing so, an ETT is passed exactly at $\mu
   =\varepsilon_c=-4t^\prime$, where the Fermi line touches the
   zone boundaries, and assumes the asteroid-like shape depicted in
   Fig.~\ref{fig:FS}.
It has been shown that such a critical value for the Fermi energy is
   stable against the RG flow for any repulsive electron-electron
   interaction \cite{Gonzalez:96,Gonzalez:01}.
Such a result has been recently confirmed also when self-energy
   effects to the quasiparticle dispersion relation are included, thus 
   demonstrating that the pinning of the Fermi surface to a Van~Hove
   singularity can actually take place for a rather wide range of hole 
   concentration \cite{Irkhin:01b}.
The condition $\mu = \varepsilon_c$ corresponds to having a saddle
   point at $X=(\pi ,0)$ in the 
   single-particle dispersion relation $\varepsilon_\bk$,
   which, for small wavevector displacements from $X$ and symmetry
   related points, can be expanded as 
\begin{equation}
\varepsilon_\bk -\varepsilon_c\sim \frac{p_{1}^{2}}{2m_{1}}-%
\frac{p_{2}^{2}}{2m_{2}} \equiv \epsilon_\bp,  
\label{eq:hyperbola}
\end{equation}
where $p_1=k_x$, $p_2=k_y-\pi $, and $m_{1,2}=[2t(1\pm 2r)]^{-1}$ are
   the eigenvalues of the effective mass tensor \cite{Morse}. 
Here and below, we choose units such that $\hbar $ and the lattice
   spacings are set equal to unity.

\begin{figure}[ht]
\centering
\includegraphics[height=0.9\columnwidth,angle=-90]{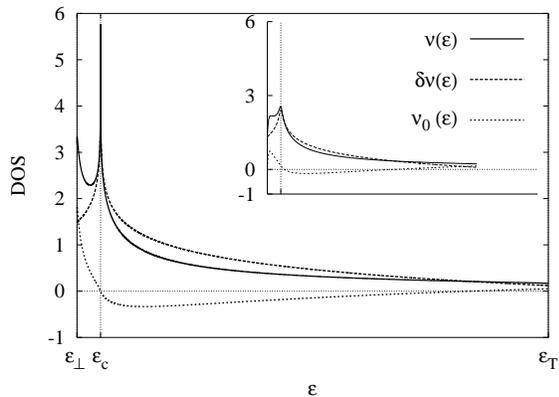}
\caption{Total density of states ($\nu$), Eq.~(\protect\ref{eq:DOSexact}), and
   regular ($\nu_0$) and singular ($\delta\nu$) contributions to the DOS,
   Eq.~(\protect\ref{eq:DOScontribs}), as a function of energy
   $\varepsilon$, ranging from the bottom ($\varepsilon_\bot$) to the
   top ($\varepsilon_\top$) of the band.
The inset shows the effect of a non-zero energy broadening
   $\Gamma$ (here, $\Gamma\sim 0.5$~\% of the total bandwidth) on the
   same quantities [see Eq.~(\protect\ref{eq:dnuG})].
}
\label{fig:DOS}
\end{figure}

For $\varepsilon_\perp \leq \varepsilon \leq \varepsilon_\top$,
   the density of states $\nu (\varepsilon )=\oint_{\varepsilon_\bk
   =\varepsilon }d\Omega_\bk |\nabla_\bk
   \varepsilon_\bk |^{-1}$ can be computed
   analytically as \cite{Xing:91}
\begin{equation}
\nu (\varepsilon )=\frac{1}{\pi
   ^{2}}\frac{1}{\sqrt{4t^{2}-\varepsilon_c \varepsilon }}K\left[
   \frac{1}{2}\sqrt{\frac{16t^{2}-(\varepsilon +\varepsilon_c
   )^{2}}{4t^{2}-\varepsilon_c \varepsilon }}\right] , 
\label{eq:DOSexact}
\end{equation}
where $K(k)$ denotes the complete elliptic integral of first kind of
   modulus $k$ (Ref.~\onlinecite{GR}).
At $\varepsilon = \varepsilon_c$, $\nu (\varepsilon )$ diverges
   logarithmically.
Making use of the appropriate asymptotic expansion for $K(k)$
   (Ref.~\onlinecite{AS}), it is then customary to identify a regular
   and a singular contribution to $\nu (\varepsilon )$ as
\begin{equation}
\nu (\varepsilon ) = \nu_0 (\varepsilon ) + \delta \nu (\varepsilon ),
\label{eq:DOScontribs}
\end{equation}
with $\nu_0 (\varepsilon )$ being a continuous function of
   $\varepsilon$ over the whole bandwidth such that $\nu_0 (\varepsilon_c )
   = 0$, and
\begin{equation}
\delta\nu(\varepsilon) = 2\rho \ln
   \left| \frac{4\sqrt{2}/\pi^2 \rho}{\varepsilon - \varepsilon_c} \right|,
\end{equation}
where $\rho^{-1} = 4 \pi^2 t\sqrt{1-4r^2}$ (Fig.~\ref{fig:DOS}).
In the following, we shall often make use of the critical
   parameter $z = \mu-\varepsilon_c$, measuring the
   distance of the chemical potential from the ETT.

\section{The effect of an ETT on the superconducting gap}
\label{sec:super}

We will now discuss the effects of the proximity to an ETT on the
   superconducting properties of a 2D, single-layer
   system, both for an $s$-wave and for a $d$-wave order parameter, in 
   the weak coupling limit.
Our starting point will be the BCS equation for the gap function
   $\Delta_\bk$, which at $T=0$ reads as
\begin{equation}
\Delta_\bk = -\frac{1}{N} \sum_{\bk^\prime} V_{\bk\bk^\prime}
   \frac{\Delta_{\bk^\prime}}{2 E_{\bk^\prime}} .
\label{eq:BCST=0}
\end{equation}
Here, $E_\bk = \sqrt{\xi_\bk^2 + |\Delta_\bk |^2}$ is the upper branch
   of the
   superconducting excitation spectrum, $V_{\bk\bk^\prime}$ denotes
   the interparticle potential, and the sum runs over all the $N$
   $\bk$ points in the 1BZ.
In the case of $s$-wave symmetry, we assume $V_{\bk\bk^\prime} = -\lambda$,
   i.e. a constant over the whole 1BZ,
   whereas in the $d$-wave case we take the potential in the separable
   form $V_{\bk\bk^\prime} = -\lambda g_\bk
   g_{\bk^\prime}$, $g_\bk = \frac{1}{2} (\cos k_x - \cos k_y )$ being
   the lowest-order lattice harmonic corresponding to $d$-wave
   symmetry.
Accordingly, one has $\Delta_\bk = \Delta_0$ in the $s$-wave case, and
   $\Delta_\bk = \Delta_0 g_\bk$ in the $d$-wave case, respectively.
It is worth emphasizing that, in both cases, the coupling constant
   $\lambda>0$ has been assumed independent of doping.
This amounts to neglecting higher-order correlation effects among
   interacting particles \cite{Perali:00}.
Moreover, the weak coupling hypothesis allows us to neglect
   renormalizations of the shape of the Fermi surface, which are
   certainly expected in the strong coupling limit, and are known to
   give rise to another kind of ETT as well \cite{Chubukov:97a}.

Eq.~(\ref{eq:BCST=0}) implicitly neglects the possibility of any
   pairing instability other than singlet superconductivity in the
   Cooper channel (characterized by a pair relative momentum ${\bf
   P}=0$).
Possible alternative intervening pairing instabilities include, \emph{e.g.,}
   antiferromagnetism and the $\pi$-triplet paired state
   \cite{Murakami:98}.
Such instabilities would be characterized by large momentum transfer
   near the `hot spots' $(0,\pi)$ and $(\pi,0)$.
They have been shown to coexist and win out singlet superconductivity
   at a mean field level near half-filling, when backward scattering
   is a relevant process \cite{Murakami:98}.
However, within our weak-coupling approximation, it is consistent to
   retain only one kind of instability (namely, Cooper pairing in the
   singlet, ${\bf P}=0$ channel, with either $s$- or $d$-wave
   symmetry), under the assumption that other instabilities are
   characterized by weaker couplings.

We will first analyze the gap equation, Eq.~(\ref{eq:BCST=0}), close
   to an ETT ($|z|\ll 4t$).
In the $s$-wave case, the summation over the 1BZ in
   Eq.~(\ref{eq:BCST=0}) can be transformed into an integral over
   energy weighted by the DOS, which we approximate by its singular
   part $\delta\nu(\varepsilon)$ in Eq.~(\ref{eq:DOScontribs}).
The Fermi line corresponding to the ETT divides the 1BZ in two
   regions, $\varepsilon_\bk <0$ and $\varepsilon_\bk >0$, which are
   electron-hole conjugated of each other.
Separating the contributions coming from these regions, the gap
   equation can be compactly written as:
\begin{equation}
\frac{1}{\lambda\rho} = S_+ + S_- ,
\label{eq:gapeqs}
\end{equation}
where $S_\pm$ represent the pairing susceptibility integrated between
   the ETT and either band edges (see App.~\ref{app:conti1} for
   details).

While in 3D BCS theory $S_\pm$ are logarithmically divergent in the limit
   $\Delta_0 \to0$ (Ref.~\onlinecite{AGD}), the proximity to an ETT in 2D makes
   them divergent as $\sim\ln^2 \Delta_0$.
Direct inspection of Eq.~(\ref{eq:gapeqs}) as well as numerical
   calculations show that $\Delta_0 (z)$ is maximum near the ETT.
Such a nonmonotonic dependence of the superconducting gap on the
   critical parameter $z$ is in agreement with the phenomenology of
   the HTSC, where $\Delta_0$ and $T_c$ are characterized by a
   parabola-like dependence on doping \cite{Zhang:93}.
This has to be contrasted with the step-like behavior observed in the 
   3D case \cite{Makarov:65}, where it has to be emphasized that no
   divergence occurs in the DOS at the ETT.
However, due to the electron-hole symmetry breaking induced by a
   non-zero hopping ratio $r$, $\Delta_0 (z)$ is not an
   even function of $z$, and its maximum will actually occur not
   exactly at $z=0$, as will be discussed below.
Solving Eq.~(\ref{eq:gapeqs}) for $\Delta_0$ in the
   weak-coupling limit ($\lambda\rho\ll 1$, $\Delta_0 \ll 4t$),
   at $z =0$ one finds:
\begin{widetext}
\begin{equation}
\Delta_0 (z=0) \simeq \frac{8\sqrt{2}}{\pi^2 \rho} \exp\left(
- \sqrt{\frac{1}{\lambda\rho} + \frac{1}{4} \ln^2 \frac{1+2r}{1-2r}}
   \right),
\qquad\qquad\mbox{($s$-wave)}
\label{eq:deltamaxs}
\end{equation}
which can be then taken as a first approximation to the gap maximum at
   $T=0$, in the $s$-wave case.
Following the same procedure, qualitatively similar results can be
   derived for the critical temperature $T_c$ as a function of
   $z$ (see also Refs.~\onlinecite{Markiewicz:97,Tsuei:90}).

In the $d$-wave case, due to the anisotropic $\bk$-dependence of the
   integrand in Eq.~(\ref{eq:BCST=0}), it is not possible to
   explicitly separate the integration over energy, and a different
   approach must be followed (see App.~\ref{app:conti1} for details).
However, the proximity to an ETT does endow the pairing susceptibility
   with an analogous asymptotic low-$\Delta_0$ behavior, like in the $s$-wave
   case, which eventually results in the following weak-coupling
   expression for the gap amplitude at $T=0$, $z=0$
   ($\lambda\rho\ll1$, $\Delta_0 \ll 4t$):
\begin{equation}
\Delta_0 (z=0) \simeq 4 t f_1 (r) \exp\left( - \sqrt{\frac{1}{\lambda\rho} +
   f_2 (r)} \right),
\qquad\qquad \mbox{($d$-wave)}
\label{eq:deltamaxd}
\end{equation}
where $f_1 (r) = 2b^{-1} (1-4r^2) (\sqrt{1+2r}+\sqrt{1-2r})^{-1}$,
   $f_2 (r) = \ln^2 (b\pi\sqrt{1-4r^2}) + 2 \ln\sqrt{1-4r^2}
   \ln[2\pi^{-1} \sqrt{1-4r^2} (\sqrt{1+2r}+\sqrt{1-2r})^{-1}]$, and
   $b=e^2 /8$.
\end{widetext}

\begin{figure}[ht]
\centering
\includegraphics[height=0.9\columnwidth,angle=-90]{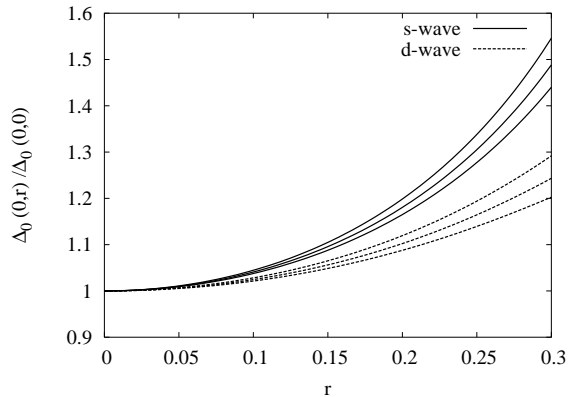}
\caption{Normalized gap amplitude $\Delta_0 (z=0,r)/\Delta_0 (z=0,r=0)$ at
   $T=0$, as a function of the hopping ratio $r=t^\prime /t$, for
   different couplings $\lambda/t = 0.9\div 1.1$ (bottom to top).
Continuous lines refer to the $s$-wave case,
   Eq.~(\protect\ref{eq:deltamaxs}), while dashed lines refer to the
   $d$-wave case, Eq.~(\protect\ref{eq:deltamaxd}).
One can recognize the direct correlation between $T_c^{\rm{max}}
   \propto \Delta_0 (z=0)$ and $r$, as observed in
   Ref.~\protect\onlinecite{Pavarini:01}.
}
\label{fig:deltamax}
\end{figure}

Fig.~\ref{fig:deltamax} shows $\Delta_0 (0)$ both in the $s$-
   and in the $d$-wave case, Eq.~(\ref{eq:deltamaxs}) and
   (\ref{eq:deltamaxd}), respectively, as a function of the hopping
   ratio $r$, for several values of $\lambda/t$.
In view of the fact that $T_c \propto \Delta_0$, as in any mean-field
   theory, Fig.~\ref{fig:deltamax} is in good qualitative agreement with
   Fig.~5 of Ref.~\onlinecite{Pavarini:01}, showing a direct correlation
   between the experimental $T_c^{\rm{max}}$ and the hopping range
   $r$ for several cuprate compounds.
Moreover, our results suggest that such an effect is a general
   consequence of the proximity to an ETT, and is roughly independent
   of the superconducting pairing symmetry.

Expanding $\Delta_0 (z)$, as implicitly defined by
   Eq.~(\ref{eq:gapeqs}) around $z=0$, one finds that the
   maximum of $\Delta_0$ actually occurs at a larger value of
   the critical parameter, which, in the $s$-wave case, is given by
\begin{equation}
z_{\rm{max}} \simeq \frac{1}{8t} \Delta_0^2 (0) \ln
   \frac{1+2r}{1-2r}.
\label{eq:zmaxs}
\end{equation}
A qualitatively analogous result applies to the $d$-wave case.
Therefore, as an effect of the electron-hole asymmetry induced by a
   non-zero hopping ratio $r$, the maximum in $\Delta_0$ is
   actually located in the hole-like region ($z_{\rm{max}} >0$),
   in agreement with the phenomenology of some hole-doped cuprate
   compounds.
For instance, a representative high-$T_c$ cuprate such as LSCO is
   characterized by an optimal doping level of $x_{\rm{opt}}
   \simeq 0.15$, lying in the hole-doped region, while a
   doping-dependent crossover from a hole-like to an electron-like FS
   has been clearly observed at a somewhat larger doping
   $x_c \simeq 0.20$ (Ref.~\onlinecite{Ino:01}).
On the contrary, no direct evidence of a change in the FS topology has
   been so far reported for Bi-2212 (Ref.~\onlinecite{Loeser:96}), whose FS
   displays a hole-like character at all dopings, including optimal
   doping.
This implies that the ETT is located at a much larger distance from optimal
   doping, which is consistent with Eq.~(\ref{eq:zmaxs}) above, given
   the larger value of the gap amplitude of Bi-2212 than that of LSCO.

\section{Effect of impurities}
\label{sec:impurities}

We now turn to consider the more realistic case, in which electron
   scattering from non-magnetic impurities is included.
Here, we will be mainly concerned with the normal state properties.
A finite quasiparticle lifetime induces a
   broadening of the energy linewidth of a quasiparticle state.
Therefore, the use of a quasiparticle description and the definition
   of a Fermi surface for impure metals can, at first sight, be
   objected.
Indeed, quasimomentum is a `good' quantum number
   only for electrons moving in a periodic potential \cite{Landau:X}.
Scattering of electrons on impurities results in momentum relaxation
   and in the corresponding smearing of the Fermi surface in momentum
   space.
The characteristic scale of such a smearing is $\sim\tau^{-1}$, where
   $\tau$ is the elastic relaxation time at low temperatures.
The value of $\tau$ can easily exceed the quasiparticle energy $\sim
   T$ even for moderate impurity concentrations.

Nevertheless, elastic scattering does not result in energy relaxation.
This means that, in principle, one can solve exactly the eigenvalue
   problem for the Hamiltonian of the electron in a lattice with some
   specific realization of the impurity potential.
The eigenstates of such Hamiltonian can then be chosen as a basis in
   the Hilbert space and the `Fermi surface' in this space can be
   defined as the surface separating the low-energy occupied
   eigenstates from the high-energy empty eigenstates at zero
   temperature.
It is evident that the Fermi surface defined in this way does exist, and
   that the elastic scattering has no effect on the quasiparticle
   lifetime in the vicinity of the Fermi surface (see also
   Refs.~\onlinecite{Varlamov:89,Blanter:94}).

The effect of a nonvanishing impurity scattering rate can then
   be accounted for in the DOS by means of a convolution between
   $\nu(\varepsilon)$ and a Lorentzian of finite width $\Gamma$:
\begin{equation}
\nu_\Gamma (\varepsilon) = \int d\xi \, \frac{1}{\pi}
   \frac{\Gamma}{(\xi-\varepsilon)^2 + \Gamma^2} \nu(\xi).
\end{equation}
Such a procedure \cite{Blanter:94} effectively smears out the
   logarithmic singularity in 
   the DOS at the ETT into a pronounced maximum of finite width
   $\sim\Gamma$ (see inset in Fig.~\ref{fig:DOS}).
Nonetheless, it is still possible to separate a `regular' and a
   `singular' contribution to $\nu_\Gamma (\varepsilon)$ as 
\begin{equation}
\nu_\Gamma (\varepsilon) = \nu_\Gamma^0 (\varepsilon) +
   \delta\nu_\Gamma (\varepsilon),
\end{equation}
with $\delta\nu_\Gamma (\varepsilon)$ now given by \cite{Blanter:94}:
\begin{equation}
\delta\nu_\Gamma (\varepsilon) = 2\rho \ln
   \frac{4\sqrt{2}/\pi^2 \rho}{\sqrt{(\varepsilon - \varepsilon_c )^2
   + \Gamma^2}} .
\label{eq:dnuG}
\end{equation}
From a physical point of view, the energy linewidth broadening
   associated to impurity scattering has the effect of `blurring' the
   Fermi line.
As a consequence, one expects that the ETT occurs slightly farther
   from $\mu=\varepsilon_c$ (i.e., for $|z|>0$),
   as soon as such a blurred Fermi line touches the border of the 1BZ
   \cite{Varlamov:89}.

\begin{figure}[ht]
\centering
\includegraphics[height=0.9\columnwidth,angle=-90]{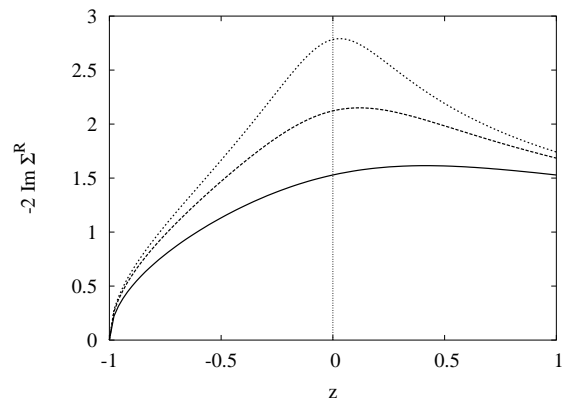}
\caption{Renormalized quasiparticle inverse lifetime
   $\tau^{-1} = -2 \Im \Sigma^{\rm R}$,
   Eq.~(\protect\ref{eq:Sigma}), resulting from isotropic impurity
   scattering.
Different curves correspond to increasing values of
   $\tau_0^{-1}$ (bottom to top).
The proximity to an ETT induces a nonmonotonic dependence of
   $\tau^{-1}$ on the critical parameter $z$, with
   $\tau^{-1}$ assuming its maximum value at $z\protect\gtrsim
   0$.
}
\label{fig:sigma}
\end{figure}

Thus far, we have assumed a constant energy linewidth $\Gamma$ over
   the whole band.
This is clearly an approximation, since the quasiparticle lifetime
   $\tau_\bk$ is generally an anisotropic quantity over the 1BZ
   \cite{Hlubina:95}.
The last statement holds true even in the simplest case of isotropic
   impurity scattering, due to the anisotropy of the single-particle
   band structure.
In particular, the proximity to an ETT in 2D endows the
   quasiparticle lifetime with a nonmonotonic behavior, in contrast to
   the 3D case, where a step-like $z$-dependence was
   found \cite{Varlamov:89}.
Following Ref.~\onlinecite{Varlamov:89}, one can write the self-consistent
   equation for the retarded quasiparticle self-energy
   $\Sigma^{\rm R}$ due to impurity scattering as
\begin{equation}
\Sigma^{\rm R} (\omega,z) = \frac{n_i |u_0 |^2}{(2\pi)^2} \int d^2 \bp \,
\left[ \epsilon_\bp + z + \omega - \Sigma^{\rm R} (\omega,z)
   \right]^{-1} ,
\end{equation}
where $\epsilon_\bp$ is the asymptotic single-particle dispersion
   relation near the saddle point defining the ETT,
   Eq.~(\ref{eq:hyperbola}), $n_i$ denotes the concentration of
   impurities, and $u_0$ is the impurity scattering strength, here
   assumed independent of $\bp$.
Performing the integrations as in
   Refs.~\onlinecite{Varlamov:85,Varlamov:89}, but now for the 2D case,
   one arrives at the self-consistent expression:
\begin{equation}
\Sigma^{\rm R} = -\frac{i}{2\tau_0} \ln \left(
\frac{\sqrt{1+z} + \sqrt{1 + \omega +
   \Sigma^{\rm R}}}{
\sqrt{-\omega-z+\Sigma^{\rm R}}}
\right),
\label{eq:Sigma}
\end{equation}
where $\tau_0^{-1} = \pi^{-1} n_i |u_0 |^2 (m_1 m_2)^{1/2}$, and all
   energies are in units of a cut-off energy $\sim\rho^{-1}$.
Fig.~\ref{fig:sigma} shows the renormalized quasiparticle inverse
   lifetime $\tau^{-1} = -2 \Im \Sigma^{\rm{R}}
   (\omega = 0,z)$ as a function of the critical
   parameter $z$, for several values of $\tau_0^{-1}$.
As anticipated, one observes a maximum in $\tau^{-1}$ at
   $z\gtrsim 0$, as an effect of crossing an ETT.

The study of a 2D superconductor in the dirty limit goes
   beyond the reach of the present analysis.
The dependence of $\Delta_0$ as well as of the DOS
   on the impurity concentration has been
   derived in the $d$-wave case in Ref.~\onlinecite{Sun:95}, for an
   isotropic dispersion relation.
In $d$-wave superconductors, Dirac-like single-particle excitations
   can be created at virtually no energy cost near the gap nodes
   \cite{Lee:97,Lee:97a,Balents:98}.
Within the QCP scenario, long-range interaction between such gapless
   modes is mediated via the fluctuations of an intervening order
   parameter at $T=0$.
Current proposals for the HTSC include the possibility of the
   proximity to a quantum ordered phase characterized by either charge 
   or (AFM) spin fluctuations, as well as fluctuations related to the
   opening of another subdominant contribution to the superconducting
   OP, usually accompanied by time-reversal breaking
   \cite{Vojta:00,Chakravarty:01}.
Recent results for 2D $d$-wave superconductors in the presence of
   disorder yield corrections to the density of states coming both
   from the diffusion (${\bf Q}=0$) and the Cooperon mode
   \cite{Khveshchenko:01}, as well as from the diffusive mode with
   ${\bf Q}=(\pi,\pi)$ \cite{Yashenkin:01}.

\section{Ginzburg-Landau stiffness near an ETT}
\label{sec:GL}

As is well known, the normal state of HTSC is characterized by several
   anomalous properties at the transition edge.
Such properties include a peak in the $c$-axis resistivity, an
   anomalously large sign-changing $c$-axis magnetoresistance, as well
   as the opening of a pseudogap, which is observed both in the
   $c$-axis optical conductivity, in tunneling experiments, and  
   in the NMR relaxation rate (see Ref.~\onlinecite{Varlamov:99} for a
   review).
On the basis of the Fermi liquid theory, it has been recently
   demonstrated that the renormalization of the one-electron DOS
   in the vicinity of the Fermi level due to the
   electron-electron interaction in the Cooper channel is able to
   explain satisfactorily many of these pseudogap-like manifestations
   both in the overdoped and in the optimally doped compounds
   \cite{Varlamov:99}.
Moving across the phase diagram of the HTSC from the overdoped,
   bad metallic region, towards underdoping, the enhancement of the
   mentioned effects correlates with an
   increase of the Ginzburg-Levanyuk parameter, ${\Gi}_{\mathrm{(2D)}}
   \approx T_c /E_{\rm{F}}$, thus making perturbation theory less reliable
   \cite{Varlamov:99}. 
Nevertheless, such a rapid growth of the normal state anomalies with
   the decrease of doping
   strongly overcomes the theoretical predictions, thus making
   it difficult to attribute such an effect to the mere shrinking of
   the FS.

Indeed, ARPES studies indicate a marked increase of the
   FS anisotropy in the $ab$-plane with underdoping, which is
   accompanied by the development of an
   extended saddle point in the electronic spectrum
   \cite{Randeria:99}. 
One can identify two characteristic energy scales related with such a
   FS, namely the size of its `bulk' part, $E_{\rm{F}} \approx 0.3$~eV, and
   the `width' of the saddle point, $|z| = |\mu - \varepsilon_c
   |\approx 0.01$~eV.
The large difference in the magnitude of such energy scales is
   therefore suggestive of a crossover, related to the special role
   played by the electronic states lying close to the saddle point
   (the so-called `slow' electrons).
Intuitively, one may expect a replacement of $E_{\rm{F}}$ with a
   small $z$ in the denominator of $\Gi_{\mathrm{(2D)}}$,
   which would result in the breakdown of the perturbative approach
   developed in Ref.~\onlinecite{Varlamov:99}.
Nevertheless, as is demonstrated below, the sole existence of an ETT in the
   electronic spectrum is not able to change the character of the isotropic
   electron-electron interaction in the Cooper channel.
This implies that the breakdown of fluctuation theory in the
   underdoped compounds has to be related to some other properties of
   the HTSC.
In order to substantiate the above statements, in the present section
   we shall derive the momentum dependence of the
   two-particle Green function in the Cooper channel near $T_c$ and
   close to an ETT.

In the case of a 2D electron system characterized by the approximate 
   spectrum given by
   Eq.~(\ref{eq:hyperbola}), close to an ETT,
   the single-particle Green function can be written as
\begin{equation}
G(\bp,\omega_n; z ) = (i\omega_n - \epsilon_\bp + z)^{-1} .
\end{equation}
Here, the electron quasi-momentum $\bp$ is measured from the saddle
   point location, as in 
   Eq.~(\ref{eq:hyperbola}), and $\omega_n =2\pi T(n+\frac{1}{2})$ are the 
   fermionic Matsubara
   frequencies. As already emphasized in Sec.~\ref{sec:model},
   the condition $z >0$
   describes the case of an open Fermi surface without any disrupted
   neck, whereas the opposite one, $z<0$, is appropriate to a closed
   Fermi surface with respect to the $\Gamma$ point
   (Fig.~\ref{fig:FS}).
The two-particle Green function in the Cooper channel $L(\bq,\Omega_{\nu}
   )$ can be expressed within the ladder approximation by means of
   the 
   polarization operator $\Pi (\bq,\Omega_{\nu} )$ as \cite{Varlamov:99}
\begin{equation}
L^{-1} (\bq,\Omega_{\nu}; z, T ) = \lambda^{-1} - \Pi (\bq,\Omega_{\nu};
      z, T ).
\label{eq:pol1}
\end{equation}
Here, $\lambda>0$ denotes the momentum independent effective
   electron-electron interaction, $\bq$ is the Cooper pair
   momentum, and 
\begin{widetext}
\begin{equation}
\Pi (\bq,\Omega_\nu ;z,T) = T \sum_{\omega_n} \int \frac{d^2
   \bp}{(2\pi)^2} G(\bp+\bq,\omega_{n+\nu}; z ) G(-\bp,-\omega_n ;z )
\equiv T \sum_{\omega_n} I (\bq,\omega_{n+\nu} , -\omega_n ;z),
\label{eq:pol2}
\end{equation}
\end{widetext}
with $\Omega_\nu = 2\pi T\nu$ the bosonic Matsubara frequencies.

The superconducting critical temperature can be characterized 
   as the temperature at which $L$ presents a pole at $\bq=0$ and
   $\Omega_\nu=0$. 
The procedure to deal with the integral in Eq.~(\ref{eq:pol2}) is
   outlined in App.~\ref{app:conti2}.
One eventually arrives at the result:
\begin{equation}
\Pi(0,0;z,T) = \frac{m}{\pi} T \sum_{\omega_n \geq
   0}^{\omega_{\rm{D}} / (2\pi 
   T)} \frac{1}{\omega_n} \ln \left( \frac{\omega_{\rm{D}}^2}
    {\omega_n^2 + z^2}
   \right),
\label{eq:pol3}
\end{equation}
where $m= \sqrt{m_1 m_2}$ is the geometric average mass around the
   saddle point, and the Debye frequency $\omega_{\rm{D}}$ has
   been introduced as a cut-off in the summation over the Matsubara
   frequencies. 
The equation for the critical temperature then reads
\begin{equation}
\lambda^{-1} = \frac{m}{\pi} T_c \sum_{\omega_n \geq 0}^{\omega_{\rm{D}} 
   / (2\pi T_c )} \frac{1}{\omega_n} \ln \left(
     \frac{\omega_{\rm{D}}^2}
       {\omega_n^2 + z^2}\right).
\end{equation}
For small $z$ ($|z|\ll \omega_{\rm{D}}$), one recovers the well-known result
   \cite{Abrikosov:93}:
\begin{equation}
T_c \sim \frac{\omega_{\rm{D}}}{2\pi} \exp \left(
   -\frac{1}{\sqrt{\lambda\rho}} \right),
\label{eq:Tcsqrt}
\end{equation}
where $\rho = m/(2\pi^2 )$ denotes the DOS at the saddle point.
Eq.~(\ref{eq:Tcsqrt}) is in agreement with the $s$-wave result for
   $\Delta_0$, Eq.~(\ref{eq:deltamaxs}), when the assumption
   of a phonon-mediated pairing mechanism is made and the limit $r\to0$ is
   taken.
Vertex and cross corrections to Eq.~(\ref{eq:Tcsqrt}) in terms of the
   Migdal adiabaticity parameter $\omega_{\mathrm{D}} /
   E_{\mathrm{F}}$ have been shown to decrease the enhancement of
   $T_c$ due to the proximity to an ETT
   \cite{Cappelluti:96a,Cappelluti:96b}.
Analogously, for the two-particle Green function close to
   the superconducting transition and in the proximity of an ETT
   ($|z|\ll T \sim T_c$), one finds
\begin{equation}
L^{-1} (0,0;z,T) = - \rho\ln\left(
   \frac{\omega_{\rm{D}}}{2\pi T_c} \right) \frac{T-T_c}{T_c}.
\end{equation}
One observes that the presence of the ETT results in the appearance of
   the additional large factor $\sim\ln(\omega_{\rm{D}}/T_c)$ in front
   of the reduced temperature.

In order to determine the superconducting stiffness tensor $\eta_i
   (z)$, one is led to consider the $\bq$-dependence of the
   polarization operator, Eq.~(\ref{eq:pol2}).
Expanding $I(\bq,\omega_n,-\omega_n;z)$ in Eq.~(\ref{eq:pol2}) for
   small $q$ and $z$, up to quadratic order in $q$, one has
\begin{widetext}
\begin{equation}
I(\bq,\omega_n, -\omega_n ; z) = I_0 (0, \omega_n, -\omega_n;z) 
+ I_1 (q^2, \omega_n, -\omega_n ;0) + I_2 (q^2, \omega_n, -\omega_n ;z),
\label{eq:pol4}
\end{equation}
\end{widetext}
where $I_j$ are defined in App.~\ref{app:conti2}, and $I_0$ has been used
   above for the definition of the critical temperature, Eq.~(\ref{eq:pol3}).
One eventually finds for the $q^2$-dependence of the two-particle
   Green function in the limit $|z|\ll T\sim T_c$ the relatively
   classical form:
\begin{equation}
L^{-1} (\bq,0;0,T) = - \rho \left[
   \ln \left(\frac{\omega_{\rm{D}}}{2\pi T_c} \right)
   \frac{T-T_c}{T_c} + \eta_1 q_1^2 + \eta_2 q_2^2
   \right],
\label{eq:L-1}
\end{equation}
where the components of the superfluid stiffness tensor are given by
\begin{equation}
\eta_i (z) = \frac{7\zeta(3)E_{\rm{F}}}{8\pi^2 T^2 m_i} ,
\end{equation}
to the lowest order in $z/E_{\rm{F}}$.
The latter expression is analogous to the result obtained
   in the standard 2D isotropic case, with $\xi_\bp = p^2 / (2m) -
   \mu$, where the superfluid stiffness reads $\eta = 7\zeta(3)
   E_{\rm{F}} / (16 \pi^2 T^2 m)$, and the DOS per spin is
   $\rho_{\mathrm{(2D)}} = m/(2\pi)$ (Ref.~\onlinecite{Varlamov:89}).
It is worth noting that in Eq.~(\ref{eq:L-1}) the effective mass of
   the fluctuating Cooper pair gets increased by a factor $\ln
   (\omega_{\rm{D}} /2\pi T_c )$, with respect to the case in which a
   parabolic spectrum is assumed.
This implies a reduced role of fluctuations near an ETT.
Indeed, our results demonstrate that the temperature range of the
   fluctuation regime is governed by essentially the same $\Gi$, while
   the propagator's effective mass is enhanced.
   
Summarizing, the results obtained above shows that a topological
   singularity in the electronic spectrum practically does not affect
   the Ginzburg-Landau stiffness, in contrast to what was
   intuitively speculated \cite{Varlamov:99}.
The reason thereof is that the value of $\eta$ is formed by gathering the
   contributions of the electronic states belonging to the whole Fermi
   surface, not only by the `slow' ones. 
Finally, we note that in the approach we followed, only the
   polarization loop Eq.~(\ref{eq:pol2}) is critical, since Cooper
   pairing of non-zero center-of-mass momentum is not taken into
   account \cite{AGD}.

\section{Conclusions}
\label{sec:conclusions}

We have reviewed the effects of an electronic topological transition
   on the superconducting properties of a 2D electron system, with an
   energy spectrum characterized by a minimum at the $\Gamma$ point
   and an extended, doping dependent saddle point at $(\pi,0)$, as is
   typical for most single-layered, hole-doped HTSC.
We analytically derived the expressions for the superconducting gap
   $\Delta_0$ at $T=0$ close to an ETT, both in the $s$-wave and in
   the $d$-wave case.
In contrast to the 3D result \cite{Makarov:65}, $\Delta_0$ turns out
   to be characterized by a nonmonotonic behavior as a function of the
   critical parameter $z$, with a maximum at $z\simeq0$, i.e. close to
   the ETT.
Due to the electron-hole symmetry-breaking induced by a non-zero value
   of the hopping ratio $r$, we find that the maximum of $\Delta_0$
   actually occurs at $z\gtrsim 0$, \emph{i.e.} it is slightly shifted
   towards the hole-like region, as observed in LSCO \cite{Ino:01}.
We point out that in previous calculations
   \cite{Tsuei:90,Pattnaik:92,Newns:92,Newns:92a,Onufrieva:00} the
   maximum of $T_c$ as a function of the critical parameter $z$ occurs 
   at the ETT, $z=0$.
This result has been often used as an objection against the relevance
   of the Van~Hove scenario for the cuprates, since ARPES data show
   that the optimal doping does not correspond to the critical point
   $z=0$, and that the FS preserves a hole-like character over the
   entire doping range for almost all hole-doped compounds (see,
   however, also the recent results of Ino \emph{et al.} for the LSCO
   compound \cite{Ino:01}).
On the contrary, we have shown that the observed difference between
   optimal doping and the doping actually corresponding to the ETT can 
   be justified by taking into account an electronic spectrum beyond
   the hyperbolic approximation.
Moreover, we find that the dependence of $\Delta_0$ on the hopping
   ratio $r$ is in good qualitative agreement with the
   phenomenological results collected by Pavarini \emph{et al.} for
   several HTSC materials \cite{Pavarini:01}, thus demonstrating the
   role of the band structure peculiarities and, in particular, of
   next-nearest neighbor hopping, in stabilizing high-temperature
   superconductivity in the cuprates.

In the presence of impurities, the Fermi line is
   effectively smeared, and one expects the anomalies due to the
   proximity to an ETT to occur at a larger value of the critical
   parameter, as soon as such a `blurred' Fermi line touches the zone
   border.
We also derived the energy dependence of the retarded quasiparticle
   self-energy $\Sigma^{\rm{R}}$ due to impurity scattering in the
   2D case, for a simplified, hyperbolic dispersion relation.
In contrast to the 3D case, $\Sigma^{\rm{R}}$ is again
   characterized by a nonmonotonic $z$-dependence, thus confirming
   that the quasiparticle lifetime $\tau_\bk$ is generally an
   anisotropic quantity over the 1BZ.

Finally, we addressed the issue of the range of fluctuations near $T_c$.
By explicitly computing the two-particle propagator in the Cooper
   channel near $T_c$, we derived the expression for the superfluid
   stiffness $\eta$ close to an ETT.
Although the Fermi velocity vanishes at the saddle point, we
   find a non-zero value of $\eta$, in complete analogy with the
   Ginzburg-Landau result for an isotropic electronic spectrum, thus
   showing that all electronic states participate in establishing the
   superconducting correlations.
Moreover, our results show that the role of fluctuations even
   diminishes near the ETT.
Indeed, while their temperature range is determined by about the same
   value of $\Gi$, the effective mass of the fluctuating Cooper pair
   increases by a factor $\ln (\omega_{\mathrm{D}} / 2 \pi T_c )$ with 
   respect to the case of a parabolic spectrum.
Therefore, in the denominator of any fluctuation contribution there
   appears a large logarithm, which implies a relative suppression of
   fluctuation effects near the ETT.

\begin{acknowledgments}
We thank J. V. Alvarez, G. Balestrino, B. Ginatempo, F. Onufrieva,
   P. Pfeuty, P. Podio-Guidugli, R. Pucci, and G. Savona for
   useful discussions.
Partial support from MURST through COFIN fund 2001 and the
   E.U. through the F.S.E. Program (G. G. N. A. and E. P.), and from
   I.N.F.M. and I.S.I. (Torino) through the `Progetto Stage'
   (G. G. N. A.) is gratefully acknowledged.
G. G. N. A. also acknowledges the D.S.T.F.E. ``Tor Vergata'' and
   P.~G. Nicosia for warm hospitality during the period in which the
   present work was brought to completion. 
\end{acknowledgments}

\appendix

\section{Evaluation of the pairing susceptibility in the
   \lowercase{$d$}-wave case} 
\label{app:conti1}

We briefly outline the derivation of the asymptotic dependence of the
   integrated pairing susceptibility close to an ETT in 2D. 
In the $s$-wave case, changing integration variables in the gap
   equation, Eq.~(\ref{eq:BCST=0}), from wavevector $\bk$ to energy,
   we can write the pairing susceptibility integrated between the ETT
   and either band edges as
\begin{equation}
S_\pm  = - \int_0^1 \frac{\ln (a_\pm
   \xi) \,d\xi}{\sqrt{(\xi \mp \zeta_\pm )^2 + \delta_\pm^2}} ,
\label{eq:pairings}
\end{equation}
where we have introduced the dimensionless auxiliary quantities
   $\zeta_\pm = z/(4t\pm 8t^\prime )$, $\delta_\pm = \Delta_0 /
   (4t\pm 8t^\prime )$, and $a_\pm = (1\pm
   2r)/(4\sqrt{2}\sqrt{1-4r^2})$.

In the $d$-wave case, the double integration over wavevector $\bk$
   cannot be reduced to a simple integration over energy, and one has
   to change variables to $\xi=\pm (\varepsilon_\bk
   -\varepsilon_c)/[4t(1\pm 2r)]$, $g=g_\bk$.
In such case, Eq.~(\ref{eq:BCST=0}) can be written as
\begin{equation}
\frac{\pi^2 t}{2\lambda} = D_+ + D_- ,
\label{eq:gapeqd}
\end{equation}
where the integrated pairing susceptibility now reads:
\begin{equation}
D_\pm = \frac{1}{4} \int_0^1 d\xi
   \int_0^{1-\xi} dg \frac{g^2}{\sqrt{(\xi\mp\zeta_\pm)^2 +
   \delta_\pm^2 g^2}} \frac{1}{\sqrt{J_1 J_2 J_3}} ,
\label{eq:pairingd}
\end{equation}
and
\begin{subequations}
\begin{eqnarray}
J_1 &&= (1-2rg)^2 + 4r[g-r+(1\pm 2r)\xi],\\
J_2 &&= (1+g)^2 - (1-\sqrt{J_1})^2/(4r^2),\\
J_3 &&= (1-g)^2 - (1-\sqrt{J_1})^2/(4r^2).
\end{eqnarray}
\end{subequations}
Eq.~(\ref{eq:pairingd}) leads to hyperelliptic integrals,
   that cannot be generally expressed in terms of known special functions
   \cite{Liu:97}.
We next set $\Omega=-(1-\sqrt{J_1})/(2r)$, with $\Omega\to\xi$ as
   $r\to0$, with which $D_+$ in Eq.~(\ref{eq:pairingd}) transforms
   into
\begin{widetext}
\begin{equation}
D_+ = \frac{1}{4} \int d\Omega \int dg \frac{g^2}{\sqrt{[r(v^2 - g^2) -
   z]^2 + \Delta_0^2 g^2}}
   \frac{1}{\sqrt{[(1+\Omega)^2 - g^2][(1-\Omega)^2 -g^2]}} ,
\label{eq:hyper}
\end{equation}
\end{widetext}
where all energies are in units of $4t$ and the integration is be to
   performed over the curvilinear triangle 
   defined by $g\geq0$, $g\leq 1-\Omega$, and $v^2 - g^2 \geq0$, with
   $v^2 = 1+\Omega^2 + \Omega/r$.
Such triangle represents the hole-like region of the 1BZ,
   $\varepsilon_\bk \geq\varepsilon_c$, in these new coordinates.

In the limit $r\to0$, one branch of the hyperbola defined by $g=v$
   reduces to the $\Omega=0$ axis, and $r (v^2 -g^2) \to\Omega$.
A further change to polar coordinates as $\Omega=\rho\cos\theta$,
   $g=\rho\sin\theta$ then allows to exactly decouple the two
   integrations in Eq.~(\ref{eq:hyper}), the integration over $\rho$
   leading to elliptic integrals.
Extracting the logarithmic divergence of these latter near
   $\theta=\pi/2$ (corresponding to the ETT, together with $\rho=1$)
   yields the answer [see Eq.~(\ref{eq:continuation}) below, with
   $r=0$]. 

In the case $r\neq0$, we could not find any such simple change of
   variables, allowing to exactly decouple the integrations in
   Eq.~(\ref{eq:hyper}).
However, since only the behavior of the dispersion relation close to
   the ETT is believed to determine the asymptotic properties of the
   integrated pairing susceptibility, we may linearly expand $r (v^2
   -g^2)$ near $\Omega=0$, $g=1$ and set $\Omega
   +2r(g-1)=\rho\cos\theta$. 
Within such approximation, Eq.~(\ref{eq:hyper}) then reads:
\begin{widetext}
\begin{equation}
D_+ = \frac{1}{4} \int_0^{\pi/2} d\theta \frac{\sin^2
   \theta}{\sqrt{\cos^2 \theta + \Delta_0^2 \sin^2 \theta}} 
   \frac{|\alpha_- \beta_+ |}{1-4r^2}
   \int_0^{\alpha_-} d\rho \frac{\rho^2}{\sqrt{(\alpha_+ +
   \rho)(\alpha_- -\rho)(\beta_+ +\rho)(\beta_- -\rho)}} ,
\label{eq:continuation}
\end{equation}
\end{widetext}
where
\begin{subequations}
\begin{eqnarray}
\alpha_\pm &&= \frac{1\pm 2r}{\cos\theta+(1-2r)\sin\theta} ,\\
\beta_\pm &&=  \frac{1\pm 2r}{\cos\theta-(1+2r)\sin\theta} .
\end{eqnarray}
\end{subequations}
Following standard methods (see, e.g., Ref.~\onlinecite{Hancock:10}),
   the inner integral can be now expressed as a combination of
   elliptic integrals, which diverge logarithmically as $\theta\to\pi/2$,
   whence Eq.~(\ref{eq:deltamaxd}) follows.

\section{Evaluation of the polarization operator}
\label{app:conti2}

The momentum integration in Eq.~(\ref{eq:pol2}) for the polarization
   operator can be performed by reducing the integration domain to the
   first quarter of the 1BZ, and dividing the latter into the two triangles
   defined by $\{ \bp \,:\, p_2 \leq (m/m_1 ) p_1 ;\,\, p_1\geq0 \}$, and $\{
   \bp \,:\, p_2 \geq (m/m_1 ) p_1 ;\,\, p_1 \geq0\}$, respectively.
One finds:
\begin{widetext}
\begin{equation}
\Pi(0,0;z,T) = T \sum_{\omega_n}I(0,\omega_n, -\omega_n;z) 
 = \frac{2m}{\pi^2} T \sum_{\omega_n} \left[ f(\omega_n, -\omega_n ;
   z) + f(\omega_n, -\omega_n ; -z) \right],
\end{equation}
where
\begin{equation}
f(\omega_n, -\omega_n; z) =  
\int_0^\infty dx_1 \int_0^x dx_2 \frac{1}{x_2^2 - x_1^2
   -i\omega_n + z} \cdot \frac{1}{x_2^2 - x_1^2
   +i\omega_n + z} .
\end{equation}
Performing the inner integration, one obtains
\begin{equation}
f(\omega_n, -\omega_n ;z) = \frac{1}{2i\omega_n}
\int_0^\infty dx \left[ \frac{\ln(x-i\sqrt{z-i\omega_n-x^2})}
{\sqrt{z-i\omega_n -x^2}}-\frac{\ln(x+i\sqrt{z-i\omega_n-x^2})}
{\sqrt{z-i\omega_n -x^2}}- \rm{H.c.} \right] .
\label{a1}
\end{equation}
\end{widetext}
The last integration can be carried out observing that 
\begin{equation}
\frac{\ln (x \pm i \sqrt{z-i\omega_n -x^2} )}{\sqrt{z-i\omega_n -x^2}}
   = \pm \frac{i}{2} \frac{d}{dx} \ln^2 ( x \pm i \sqrt{z-i\omega_n
   -x^2} ),
\end{equation}
and choosing the branches of the logarithms in order to make them
   complex conjugated of each other.
Finally, we obtain:
\begin{equation}
I(0,\omega_n, -\omega_n; z)= \frac{m}{2 \pi}\frac{1}{|\omega_n|} \ln
\frac{{\omega_{\rm{D}}^2}}{z^2 + \omega_n^2} .
\end{equation}

In order to calculate the GL stiffness, let us now expand the
   polarization operator up to quadratic order in $q$.
One has:
\begin{equation}
\Pi (q,0;z,T)= \Pi (0,0;z,T) + \rho \sum_{i=1,2} \eta_i q_i^2 ,
\end{equation}
where the components $\eta_i$ of the stiffness tensor are defined by
\begin{widetext}
\begin{equation}
\rho \sum_{i=1,2} \eta_i q_i^2
=T\sum_{\omega_n}[I_1(q^2,\omega_n,-\omega_n;0) + I_2
(q^2,\omega_n,-\omega_n ;z) + \ldots].
\end{equation}
One explicitly finds:
\begin{equation}
\eta_i (z) = \frac{1}{2m_i} T \sum_{\omega_n} \int 
\frac{d^2 \bx }{x_2^2 - x_1^2 -i\omega_n}
\cdot \frac{x_i^2}{(x_2^2 - x_1^2 + i\omega_n )^3} 
- \frac{z}{2m_i} T \sum_{\omega_n} \int d^2 \bx
\sum_{k=0}^1 \frac{3^k}{(x_2^2 -x_1^2 + i\omega_n )^{3+k}} \cdot
   \frac{x_i^2}{(x_2^2 -x_1^2 -i\omega_n )^{2-k}} 
\end{equation}
to the first order in $z/E_{\rm{F}}$.
The first integral, giving the principal contribution to the
   stiffness, can be evaluated using polar coordinates as:
\begin{eqnarray}
\eta_i (z=0) &&= \frac{T}{m_i} \sum_{\omega_n} \frac{1}{\omega_n^2}
\int_0^{2\pi} d\phi \int_0^\infty dr \frac{r^3 (r^4 \cos^2 2\phi -1)
   \cos^2 \phi}{(r^4 \cos^2 2\phi +1)^3} \nonumber \\
&&= - \frac{\pi T}{4m_i} \sum_{\omega_n} \frac{1}{\omega_n^2}
\int_0^\infty dt \left[ \frac{\partial}{\partial\alpha} +
   \frac{\partial^2}{\partial \alpha^2} \right]
   \left. \frac{1}{\sqrt{\alpha} \sqrt{t+\alpha}} \right|_{\alpha=1}
   \nonumber \\
&&\sim \frac{\pi T}{m_i} \sum_{n\geq 0} \frac{1}{\omega_n^3} E_{\rm{F}} =
\frac{7\zeta(3)E_{\rm{F}}}{8\pi^2 T^2 m_i} .
\end{eqnarray}
\end{widetext}

\bibliographystyle{apsrev}
\bibliography{a,b,c,d,e,f,g,h,i,j,k,l,m,n,o,p,q,r,s,t,u,v,w,x,y,z,zzproceedings,Angilella,notes}

\begin{thebibliography}{76}
\expandafter\ifx\csname natexlab\endcsname\relax\def\natexlab#1{#1}\fi
\expandafter\ifx\csname bibnamefont\endcsname\relax
  \def\bibnamefont#1{#1}\fi
\expandafter\ifx\csname bibfnamefont\endcsname\relax
  \def\bibfnamefont#1{#1}\fi
\expandafter\ifx\csname citenamefont\endcsname\relax
  \def\citenamefont#1{#1}\fi
\expandafter\ifx\csname url\endcsname\relax
  \def\url#1{\texttt{#1}}\fi
\expandafter\ifx\csname urlprefix\endcsname\relax\def\urlprefix{URL }\fi
\providecommand{\bibinfo}[2]{#2}
\providecommand{\eprint}[2][]{\url{#2}}

\bibitem[{\citenamefont{Anderson}(1997)}]{Anderson:97}
\bibinfo{author}{\bibfnamefont{P.~W.} \bibnamefont{Anderson}},
  \emph{\bibinfo{title}{The Theory of Superconductivity in the High-{$T_c$}
  Cuprates}} (\bibinfo{publisher}{Princeton University Press},
  \bibinfo{address}{Princeton NJ}, \bibinfo{year}{1997}).

\bibitem[{\citenamefont{Clarke and Strong}(1997)}]{Clarke:97}
\bibinfo{author}{\bibfnamefont{D.~G.} \bibnamefont{Clarke}} \bibnamefont{and}
  \bibinfo{author}{\bibfnamefont{S.~P.} \bibnamefont{Strong}},
  \bibinfo{journal}{Adv. Phys.} \textbf{\bibinfo{volume}{46}},
  \bibinfo{pages}{545} (\bibinfo{year}{1997}).

\bibitem[{\citenamefont{Shen and Dessau}(1995)}]{Shen:95}
\bibinfo{author}{\bibfnamefont{Z.-X.} \bibnamefont{Shen}} \bibnamefont{and}
  \bibinfo{author}{\bibfnamefont{D.~S.} \bibnamefont{Dessau}},
  \bibinfo{journal}{Phys. Rep.} \textbf{\bibinfo{volume}{253}},
  \bibinfo{pages}{1} (\bibinfo{year}{1995}).

\bibitem[{\citenamefont{{McKenzie}}(1997)}]{McKenzie:97}
\bibinfo{author}{\bibfnamefont{R.~H.} \bibnamefont{{McKenzie}}},
  \bibinfo{journal}{Science} \textbf{\bibinfo{volume}{278}},
  \bibinfo{pages}{820} (\bibinfo{year}{1997}).

\bibitem[{\citenamefont{Yokoya et~al.}(1996)\citenamefont{Yokoya, Chainani,
  Takahashi, {Katayama-Yoshida}, Kasai, and Tokura}}]{Yokoya:96}
\bibinfo{author}{\bibfnamefont{T.}~\bibnamefont{Yokoya}},
  \bibinfo{author}{\bibfnamefont{A.}~\bibnamefont{Chainani}},
  \bibinfo{author}{\bibfnamefont{T.}~\bibnamefont{Takahashi}},
  \bibinfo{author}{\bibfnamefont{H.}~\bibnamefont{{Katayama-Yoshida}}},
  \bibinfo{author}{\bibfnamefont{M.}~\bibnamefont{Kasai}}, \bibnamefont{and}
  \bibinfo{author}{\bibfnamefont{Y.}~\bibnamefont{Tokura}},
  \bibinfo{journal}{Phys. Rev. Lett.} \textbf{\bibinfo{volume}{76}},
  \bibinfo{pages}{3009} (\bibinfo{year}{1996}), \bibinfo{note}{see also A. P.
  Mackenzie, S. R. Julian, G. G. Lonzarich, Y. Maeno and T. Fujita, Phys. Rev.
  Lett. {\bf 78}, 2271 (1997), and T. Yokoya \emph{et al., ibid.,} p. 2272.}

\bibitem[{\citenamefont{Ino et~al.}(2002)\citenamefont{Ino, Kim, Nakamura,
  Yoshida, Mizokawa, Shen, Fujimori, Kakeshita, Eisaki, and Uchida}}]{Ino:01}
\bibinfo{author}{\bibfnamefont{A.}~\bibnamefont{Ino}},
  \bibinfo{author}{\bibfnamefont{C.}~\bibnamefont{Kim}},
  \bibinfo{author}{\bibfnamefont{M.}~\bibnamefont{Nakamura}},
  \bibinfo{author}{\bibfnamefont{T.}~\bibnamefont{Yoshida}},
  \bibinfo{author}{\bibfnamefont{T.}~\bibnamefont{Mizokawa}},
  \bibinfo{author}{\bibfnamefont{Z.}~\bibnamefont{Shen}},
  \bibinfo{author}{\bibfnamefont{A.}~\bibnamefont{Fujimori}},
  \bibinfo{author}{\bibfnamefont{T.}~\bibnamefont{Kakeshita}},
  \bibinfo{author}{\bibfnamefont{H.}~\bibnamefont{Eisaki}}, \bibnamefont{and}
  \bibinfo{author}{\bibfnamefont{S.}~\bibnamefont{Uchida}},
  \bibinfo{journal}{Phys. Rev. B} \textbf{\bibinfo{volume}{65}},
  \bibinfo{pages}{094504} (\bibinfo{year}{2002}).

\bibitem[{\citenamefont{Markiewicz}(1997)}]{Markiewicz:97}
\bibinfo{author}{\bibfnamefont{R.~S.} \bibnamefont{Markiewicz}},
  \bibinfo{journal}{J. Phys. Chem. Solids} \textbf{\bibinfo{volume}{58}},
  \bibinfo{pages}{1179} (\bibinfo{year}{1997}).

\bibitem[{not({\natexlab{a}})}]{note:RBCO}
\bibinfo{note}{The generic relevance of the Van~Hove scenario for high-$T_c$
  superconductivity in the cuprates has been however questioned at least for
  the RBCO family [J. L. Tallon \emph{et al.,} Phys. Rev. B {\bf 53}, R11972
  (1996)]. Yet another open issue is the competition of an ETT with an
  intervening structural instability, as exhibited by LSCO near optimal doping,
  and with the tendency exhibited by some layered cuprates to self-dope.}

\bibitem[{\citenamefont{Varlamov et~al.}(1989)\citenamefont{Varlamov, Egorov,
  and Pantsulaya}}]{Varlamov:89}
\bibinfo{author}{\bibfnamefont{A.~A.} \bibnamefont{Varlamov}},
  \bibinfo{author}{\bibfnamefont{V.~S.} \bibnamefont{Egorov}},
  \bibnamefont{and} \bibinfo{author}{\bibfnamefont{A.~V.}
  \bibnamefont{Pantsulaya}}, \bibinfo{journal}{Adv. Phys.}
  \textbf{\bibinfo{volume}{38}}, \bibinfo{pages}{469} (\bibinfo{year}{1989}).

\bibitem[{\citenamefont{Lifshitz}(1960)}]{Lifshitz:60}
\bibinfo{author}{\bibfnamefont{I.~M.} \bibnamefont{Lifshitz}},
  \bibinfo{journal}{Sov. Phys. JETP} \textbf{\bibinfo{volume}{11}},
  \bibinfo{pages}{1130} (\bibinfo{year}{1960}), \bibinfo{note}{[Zh. Eksp. Teor.
  Fiz. {\bf 38}, 1569 (1960)]}.

\bibitem[{\citenamefont{Dorbolo et~al.}(1998)\citenamefont{Dorbolo, Ausloos,
  and Houssa}}]{Dorbolo:98}
\bibinfo{author}{\bibfnamefont{S.}~\bibnamefont{Dorbolo}},
  \bibinfo{author}{\bibfnamefont{M.}~\bibnamefont{Ausloos}}, \bibnamefont{and}
  \bibinfo{author}{\bibfnamefont{M.}~\bibnamefont{Houssa}},
  \bibinfo{journal}{Phys. Rev. B} \textbf{\bibinfo{volume}{57}},
  \bibinfo{pages}{5401} (\bibinfo{year}{1998}).

\bibitem[{\citenamefont{Blanter et~al.}(1994)\citenamefont{Blanter, Kaganov,
  Pantsulaya, and Varlamov}}]{Blanter:94}
\bibinfo{author}{\bibfnamefont{Y.~M.} \bibnamefont{Blanter}},
  \bibinfo{author}{\bibfnamefont{M.~I.} \bibnamefont{Kaganov}},
  \bibinfo{author}{\bibfnamefont{A.~V.} \bibnamefont{Pantsulaya}},
  \bibnamefont{and} \bibinfo{author}{\bibfnamefont{A.~A.}
  \bibnamefont{Varlamov}}, \bibinfo{journal}{Phys. Rep.}
  \textbf{\bibinfo{volume}{245}}, \bibinfo{pages}{159} (\bibinfo{year}{1994}).

\bibitem[{\citenamefont{Bruno et~al.}(1994)\citenamefont{Bruno, Ginatempo,
  Giuliano, Ruban, and Velikov}}]{Bruno:94}
\bibinfo{author}{\bibfnamefont{E.}~\bibnamefont{Bruno}},
  \bibinfo{author}{\bibfnamefont{B.}~\bibnamefont{Ginatempo}},
  \bibinfo{author}{\bibfnamefont{E.~S.} \bibnamefont{Giuliano}},
  \bibinfo{author}{\bibfnamefont{A.~V.} \bibnamefont{Ruban}}, \bibnamefont{and}
  \bibinfo{author}{\bibfnamefont{Y.~K.} \bibnamefont{Velikov}},
  \bibinfo{journal}{Phys. Rep.} \textbf{\bibinfo{volume}{249}},
  \bibinfo{pages}{353} (\bibinfo{year}{1994}).

\bibitem[{not({\natexlab{b}})}]{note:fluctuations}
\bibinfo{note}{It should be noted that quantum critical fluctuations may give
  rise to a strong DOS renormalization, as has been shown in the case of a
  disordered electron system near a ferromagnetic quantum critical point at
  $T=0$ [D. Belitz \protect\emph{et al.,} Phys. Rev. Lett. {\bf 85}, 4602
  (2000)]. In the following, however, we shall neglect such effects, for the
  sake of simplicity.}

\bibitem[{\citenamefont{Tsuei et~al.}(1990)\citenamefont{Tsuei, Newns, Chi, and
  Pattnaik}}]{Tsuei:90}
\bibinfo{author}{\bibfnamefont{C.~C.} \bibnamefont{Tsuei}},
  \bibinfo{author}{\bibfnamefont{D.~M.} \bibnamefont{Newns}},
  \bibinfo{author}{\bibfnamefont{C.~C.} \bibnamefont{Chi}}, \bibnamefont{and}
  \bibinfo{author}{\bibfnamefont{P.~C.} \bibnamefont{Pattnaik}},
  \bibinfo{journal}{Phys. Rev. Lett.} \textbf{\bibinfo{volume}{65}},
  \bibinfo{pages}{2724} (\bibinfo{year}{1990}).

\bibitem[{\citenamefont{Zhang and Sato}(1993)}]{Zhang:93}
\bibinfo{author}{\bibfnamefont{H.}~\bibnamefont{Zhang}} \bibnamefont{and}
  \bibinfo{author}{\bibfnamefont{H.}~\bibnamefont{Sato}},
  \bibinfo{journal}{Phys. Rev. Lett.} \textbf{\bibinfo{volume}{70}},
  \bibinfo{pages}{1697} (\bibinfo{year}{1993}).

\bibitem[{\citenamefont{Wijngaarden et~al.}(1999)\citenamefont{Wijngaarden,
  Jover, and Griessen}}]{Wijngaarden:99}
\bibinfo{author}{\bibfnamefont{R.~J.} \bibnamefont{Wijngaarden}},
  \bibinfo{author}{\bibfnamefont{D.~T.} \bibnamefont{Jover}}, \bibnamefont{and}
  \bibinfo{author}{\bibfnamefont{R.}~\bibnamefont{Griessen}},
  \bibinfo{journal}{Physica B} \textbf{\bibinfo{volume}{265}},
  \bibinfo{pages}{128} (\bibinfo{year}{1999}).

\bibitem[{\citenamefont{Makarov and {Bar'yakhtar}}(1965)}]{Makarov:65}
\bibinfo{author}{\bibfnamefont{V.~I.} \bibnamefont{Makarov}} \bibnamefont{and}
  \bibinfo{author}{\bibfnamefont{V.~G.} \bibnamefont{{Bar'yakhtar}}},
  \bibinfo{journal}{Sov. Phys. JETP} \textbf{\bibinfo{volume}{21}},
  \bibinfo{pages}{1151} (\bibinfo{year}{1965}), \bibinfo{note}{[Zh. Eksp. Teor.
  Fiz. {\bf 48}, 1717 (1965)]}.

\bibitem[{\citenamefont{Pattnaik et~al.}(1992)\citenamefont{Pattnaik, Kane,
  Newns, and Tsuei}}]{Pattnaik:92}
\bibinfo{author}{\bibfnamefont{P.~C.} \bibnamefont{Pattnaik}},
  \bibinfo{author}{\bibfnamefont{C.~L.} \bibnamefont{Kane}},
  \bibinfo{author}{\bibfnamefont{D.~M.} \bibnamefont{Newns}}, \bibnamefont{and}
  \bibinfo{author}{\bibfnamefont{C.~C.} \bibnamefont{Tsuei}},
  \bibinfo{journal}{Phys. Rev. B} \textbf{\bibinfo{volume}{45}},
  \bibinfo{pages}{5714} (\bibinfo{year}{1992}).

\bibitem[{\citenamefont{Newns et~al.}(1992{\natexlab{a}})\citenamefont{Newns,
  Krishnamurthy, Pattnaik, Tsuei, and Kane}}]{Newns:92}
\bibinfo{author}{\bibfnamefont{D.~M.} \bibnamefont{Newns}},
  \bibinfo{author}{\bibfnamefont{H.~R.} \bibnamefont{Krishnamurthy}},
  \bibinfo{author}{\bibfnamefont{P.~C.} \bibnamefont{Pattnaik}},
  \bibinfo{author}{\bibfnamefont{C.~C.} \bibnamefont{Tsuei}}, \bibnamefont{and}
  \bibinfo{author}{\bibfnamefont{C.~L.} \bibnamefont{Kane}},
  \bibinfo{journal}{Phys. Rev. Lett.} \textbf{\bibinfo{volume}{69}},
  \bibinfo{pages}{1264} (\bibinfo{year}{1992}{\natexlab{a}}).

\bibitem[{\citenamefont{Newns et~al.}(1992{\natexlab{b}})\citenamefont{Newns,
  Tsuei, and Pattnaik}}]{Newns:92a}
\bibinfo{author}{\bibfnamefont{D.~M.} \bibnamefont{Newns}},
  \bibinfo{author}{\bibfnamefont{C.~C.} \bibnamefont{Tsuei}}, \bibnamefont{and}
  \bibinfo{author}{\bibfnamefont{P.~C.} \bibnamefont{Pattnaik}},
  \bibinfo{journal}{Phys. Rev. B} \textbf{\bibinfo{volume}{52}},
  \bibinfo{pages}{13611} (\bibinfo{year}{1992}{\natexlab{b}}).

\bibitem[{\citenamefont{Gopalan et~al.}(1992)\citenamefont{Gopalan, Gunnarsson,
  and Andersen}}]{Gopalan:92}
\bibinfo{author}{\bibfnamefont{S.}~\bibnamefont{Gopalan}},
  \bibinfo{author}{\bibfnamefont{O.}~\bibnamefont{Gunnarsson}},
  \bibnamefont{and} \bibinfo{author}{\bibfnamefont{O.~K.}
  \bibnamefont{Andersen}}, \bibinfo{journal}{Phys. Rev. B}
  \textbf{\bibinfo{volume}{46}}, \bibinfo{pages}{11798} (\bibinfo{year}{1992}).

\bibitem[{\citenamefont{Dzyaloshinskii}(1996)}]{Dzyaloshinskii:96}
\bibinfo{author}{\bibfnamefont{I.}~\bibnamefont{Dzyaloshinskii}},
  \bibinfo{journal}{J. Phys. I} \textbf{\bibinfo{volume}{6}},
  \bibinfo{pages}{119} (\bibinfo{year}{1996}).

\bibitem[{\citenamefont{Chakravarty et~al.}(2001)\citenamefont{Chakravarty,
  Laughlin, Morr, and Nayak}}]{Chakravarty:01}
\bibinfo{author}{\bibfnamefont{S.}~\bibnamefont{Chakravarty}},
  \bibinfo{author}{\bibfnamefont{R.~B.} \bibnamefont{Laughlin}},
  \bibinfo{author}{\bibfnamefont{D.~K.} \bibnamefont{Morr}}, \bibnamefont{and}
  \bibinfo{author}{\bibfnamefont{C.}~\bibnamefont{Nayak}},
  \bibinfo{journal}{Phys. Rev. B} \textbf{\bibinfo{volume}{63}},
  \bibinfo{pages}{094503} (\bibinfo{year}{2001}).

\bibitem[{\citenamefont{Vojta et~al.}(2000)\citenamefont{Vojta, Zhang, and
  Sachdev}}]{Vojta:00}
\bibinfo{author}{\bibfnamefont{M.}~\bibnamefont{Vojta}},
  \bibinfo{author}{\bibfnamefont{Y.}~\bibnamefont{Zhang}}, \bibnamefont{and}
  \bibinfo{author}{\bibfnamefont{S.}~\bibnamefont{Sachdev}},
  \bibinfo{journal}{Phys. Rev. Lett.} \textbf{\bibinfo{volume}{85}},
  \bibinfo{pages}{4940} (\bibinfo{year}{2000}).

\bibitem[{\citenamefont{Onufrieva et~al.}(1999)\citenamefont{Onufrieva, Pfeuty,
  and Kiselev}}]{Onufrieva:99a}
\bibinfo{author}{\bibfnamefont{F.}~\bibnamefont{Onufrieva}},
  \bibinfo{author}{\bibfnamefont{P.}~\bibnamefont{Pfeuty}}, \bibnamefont{and}
  \bibinfo{author}{\bibfnamefont{M.}~\bibnamefont{Kiselev}},
  \bibinfo{journal}{Phys. Rev. Lett.} \textbf{\bibinfo{volume}{82}},
  \bibinfo{pages}{2370} (\bibinfo{year}{1999}).

\bibitem[{\citenamefont{Onufrieva and Pfeuty}(1999)}]{Onufrieva:99b}
\bibinfo{author}{\bibfnamefont{F.}~\bibnamefont{Onufrieva}} \bibnamefont{and}
  \bibinfo{author}{\bibfnamefont{P.}~\bibnamefont{Pfeuty}},
  \bibinfo{journal}{Phys. Rev. Lett.} \textbf{\bibinfo{volume}{82}},
  \bibinfo{pages}{3136} (\bibinfo{year}{1999}), \bibinfo{note}{[Phys. Rev.
  Lett. {\bf 83}, 1271 (1999)]}.

\bibitem[{\citenamefont{Onufrieva and Pfeuty}(2000)}]{Onufrieva:00}
\bibinfo{author}{\bibfnamefont{F.}~\bibnamefont{Onufrieva}} \bibnamefont{and}
  \bibinfo{author}{\bibfnamefont{P.}~\bibnamefont{Pfeuty}},
  \bibinfo{journal}{Phys. Rev. B} \textbf{\bibinfo{volume}{61}},
  \bibinfo{pages}{799} (\bibinfo{year}{2000}).

\bibitem[{\citenamefont{Onufrieva et~al.}(1996)\citenamefont{Onufrieva, Petit,
  and Sidis}}]{Onufrieva:96}
\bibinfo{author}{\bibfnamefont{F.}~\bibnamefont{Onufrieva}},
  \bibinfo{author}{\bibfnamefont{S.}~\bibnamefont{Petit}}, \bibnamefont{and}
  \bibinfo{author}{\bibfnamefont{Y.}~\bibnamefont{Sidis}},
  \bibinfo{journal}{Phys. Rev. B} \textbf{\bibinfo{volume}{54}},
  \bibinfo{pages}{12464} (\bibinfo{year}{1996}).

\bibitem[{\citenamefont{Murakami and Fukuyama}(1998)}]{Murakami:98}
\bibinfo{author}{\bibfnamefont{M.}~\bibnamefont{Murakami}} \bibnamefont{and}
  \bibinfo{author}{\bibfnamefont{H.}~\bibnamefont{Fukuyama}},
  \bibinfo{journal}{J. Phys. Soc. Japan} \textbf{\bibinfo{volume}{67}},
  \bibinfo{pages}{41} (\bibinfo{year}{1998}).

\bibitem[{\citenamefont{Alvarez et~al.}(1998)\citenamefont{Alvarez,
  {Gonz\'alez}, Guinea, and Vozmediano}}]{Alvarez:98a}
\bibinfo{author}{\bibfnamefont{J.~V.} \bibnamefont{Alvarez}},
  \bibinfo{author}{\bibfnamefont{J.}~\bibnamefont{{Gonz\'alez}}},
  \bibinfo{author}{\bibfnamefont{F.}~\bibnamefont{Guinea}}, \bibnamefont{and}
  \bibinfo{author}{\bibfnamefont{M.~A.~H.} \bibnamefont{Vozmediano}},
  \bibinfo{journal}{J. Phys. Soc. Japan} \textbf{\bibinfo{volume}{67}},
  \bibinfo{pages}{1868} (\bibinfo{year}{1998}).

\bibitem[{\citenamefont{{Gonz\'alez} et~al.}(2000)\citenamefont{{Gonz\'alez},
  Guinea, and Vozmediano}}]{Gonzalez:00}
\bibinfo{author}{\bibfnamefont{J.}~\bibnamefont{{Gonz\'alez}}},
  \bibinfo{author}{\bibfnamefont{F.}~\bibnamefont{Guinea}}, \bibnamefont{and}
  \bibinfo{author}{\bibfnamefont{M.~A.~H.} \bibnamefont{Vozmediano}},
  \bibinfo{journal}{Phys. Rev. Lett.} \textbf{\bibinfo{volume}{84}},
  \bibinfo{pages}{4930} (\bibinfo{year}{2000}).

\bibitem[{\citenamefont{Alvarez and Gonz{\'a}lez}(1998)}]{Alvarez:98}
\bibinfo{author}{\bibfnamefont{J.~V.} \bibnamefont{Alvarez}} \bibnamefont{and}
  \bibinfo{author}{\bibfnamefont{J.}~\bibnamefont{Gonz{\'a}lez}},
  \bibinfo{journal}{Europhys. Lett.} \textbf{\bibinfo{volume}{44}},
  \bibinfo{pages}{641} (\bibinfo{year}{1998}).

\bibitem[{\citenamefont{Irkhin et~al.}(2001{\natexlab{a}})\citenamefont{Irkhin,
  Katanin, and Katsnelson}}]{Irkhin:01}
\bibinfo{author}{\bibfnamefont{V.}~\bibnamefont{Irkhin}},
  \bibinfo{author}{\bibfnamefont{A.~A.} \bibnamefont{Katanin}},
  \bibnamefont{and} \bibinfo{author}{\bibfnamefont{M.~I.}
  \bibnamefont{Katsnelson}}, \bibinfo{journal}{Phys. Rev. B}
  \textbf{\bibinfo{volume}{64}}, \bibinfo{pages}{165107}
  (\bibinfo{year}{2001}{\natexlab{a}}).

\bibitem[{\citenamefont{Kiselev et~al.}(2000)\citenamefont{Kiselev, Bouis,
  Onufrieva, and Pfeuty}}]{Kiselev:00}
\bibinfo{author}{\bibfnamefont{M.}~\bibnamefont{Kiselev}},
  \bibinfo{author}{\bibfnamefont{F.}~\bibnamefont{Bouis}},
  \bibinfo{author}{\bibfnamefont{F.}~\bibnamefont{Onufrieva}},
  \bibnamefont{and} \bibinfo{author}{\bibfnamefont{P.}~\bibnamefont{Pfeuty}},
  \bibinfo{journal}{Eur. Phys. J. B} \textbf{\bibinfo{volume}{16}},
  \bibinfo{pages}{601} (\bibinfo{year}{2000}).

\bibitem[{\citenamefont{Honerkamp et~al.}(2001)\citenamefont{Honerkamp,
  Salmhofer, Furukawa, and Rice}}]{Honerkamp:01}
\bibinfo{author}{\bibfnamefont{C.}~\bibnamefont{Honerkamp}},
  \bibinfo{author}{\bibfnamefont{M.}~\bibnamefont{Salmhofer}},
  \bibinfo{author}{\bibfnamefont{N.}~\bibnamefont{Furukawa}}, \bibnamefont{and}
  \bibinfo{author}{\bibfnamefont{T.~M.} \bibnamefont{Rice}},
  \bibinfo{journal}{Phys. Rev. B} \textbf{\bibinfo{volume}{63}},
  \bibinfo{pages}{035109} (\bibinfo{year}{2001}).

\bibitem[{\citenamefont{Furukawa et~al.}(1998)\citenamefont{Furukawa, Rice, and
  Salmhofer}}]{Furukawa:98}
\bibinfo{author}{\bibfnamefont{N.}~\bibnamefont{Furukawa}},
  \bibinfo{author}{\bibfnamefont{T.~M.} \bibnamefont{Rice}}, \bibnamefont{and}
  \bibinfo{author}{\bibfnamefont{M.}~\bibnamefont{Salmhofer}},
  \bibinfo{journal}{Phys. Rev. Lett.} \textbf{\bibinfo{volume}{81}},
  \bibinfo{pages}{3195} (\bibinfo{year}{1998}).

\bibitem[{\citenamefont{Gabovich et~al.}(2001)\citenamefont{Gabovich, Voitenko,
  Annett, and Ausloos}}]{Gabovich:01}
\bibinfo{author}{\bibfnamefont{A.~M.} \bibnamefont{Gabovich}},
  \bibinfo{author}{\bibfnamefont{A.~I.} \bibnamefont{Voitenko}},
  \bibinfo{author}{\bibfnamefont{J.~F.} \bibnamefont{Annett}},
  \bibnamefont{and} \bibinfo{author}{\bibfnamefont{M.}~\bibnamefont{Ausloos}},
  \bibinfo{journal}{Supercond. Sci. Technol.} \textbf{\bibinfo{volume}{14}},
  \bibinfo{pages}{R1} (\bibinfo{year}{2001}).

\bibitem[{\citenamefont{Pavarini et~al.}(2001)\citenamefont{Pavarini, Dasgupta,
  {Saha-Dasgupta}, Jepsen, and Andersen}}]{Pavarini:01}
\bibinfo{author}{\bibfnamefont{E.}~\bibnamefont{Pavarini}},
  \bibinfo{author}{\bibfnamefont{I.}~\bibnamefont{Dasgupta}},
  \bibinfo{author}{\bibfnamefont{T.}~\bibnamefont{{Saha-Dasgupta}}},
  \bibinfo{author}{\bibfnamefont{O.}~\bibnamefont{Jepsen}}, \bibnamefont{and}
  \bibinfo{author}{\bibfnamefont{O.~K.} \bibnamefont{Andersen}},
  \bibinfo{journal}{Phys. Rev. Lett.} \textbf{\bibinfo{volume}{87}},
  \bibinfo{pages}{047003} (\bibinfo{year}{2001}).

\bibitem[{\citenamefont{Varlamov et~al.}(1999)\citenamefont{Varlamov,
  Balestrino, Milani, and Livanov}}]{Varlamov:99}
\bibinfo{author}{\bibfnamefont{A.~A.} \bibnamefont{Varlamov}},
  \bibinfo{author}{\bibfnamefont{G.}~\bibnamefont{Balestrino}},
  \bibinfo{author}{\bibfnamefont{E.}~\bibnamefont{Milani}}, \bibnamefont{and}
  \bibinfo{author}{\bibfnamefont{D.~V.} \bibnamefont{Livanov}},
  \bibinfo{journal}{Adv. Phys.} \textbf{\bibinfo{volume}{48}},
  \bibinfo{pages}{655} (\bibinfo{year}{1999}).

\bibitem[{\citenamefont{Perali et~al.}(2000)\citenamefont{Perali, Castellani,
  {Di Castro}, Grilli, Piegari, and Varlamov}}]{Perali:00}
\bibinfo{author}{\bibfnamefont{A.}~\bibnamefont{Perali}},
  \bibinfo{author}{\bibfnamefont{C.}~\bibnamefont{Castellani}},
  \bibinfo{author}{\bibfnamefont{C.}~\bibnamefont{{Di Castro}}},
  \bibinfo{author}{\bibfnamefont{M.}~\bibnamefont{Grilli}},
  \bibinfo{author}{\bibfnamefont{E.}~\bibnamefont{Piegari}}, \bibnamefont{and}
  \bibinfo{author}{\bibfnamefont{A.~A.} \bibnamefont{Varlamov}},
  \bibinfo{journal}{Phys. Rev. B} \textbf{\bibinfo{volume}{62}},
  \bibinfo{pages}{R9295} (\bibinfo{year}{2000}).

\bibitem[{\citenamefont{Andersen et~al.}(1995)\citenamefont{Andersen,
  Liechtenstein, Jepsen, and Paulsen}}]{Andersen:95}
\bibinfo{author}{\bibfnamefont{O.~K.} \bibnamefont{Andersen}},
  \bibinfo{author}{\bibfnamefont{A.~I.} \bibnamefont{Liechtenstein}},
  \bibinfo{author}{\bibfnamefont{O.}~\bibnamefont{Jepsen}}, \bibnamefont{and}
  \bibinfo{author}{\bibfnamefont{F.}~\bibnamefont{Paulsen}},
  \bibinfo{journal}{J. Phys. Chem. Solids} \textbf{\bibinfo{volume}{56}},
  \bibinfo{pages}{1573} (\bibinfo{year}{1995}).

\bibitem[{\citenamefont{Randeria and Campuzano}(1999)}]{Randeria:99}
\bibinfo{author}{\bibfnamefont{M.}~\bibnamefont{Randeria}} \bibnamefont{and}
  \bibinfo{author}{\bibfnamefont{J.}~\bibnamefont{Campuzano}}, in
  \emph{\bibinfo{booktitle}{Models and phenomenology for conventional and
  high-temperature superconductivity}}, edited by
  \bibinfo{editor}{\bibfnamefont{G.}~\bibnamefont{Iadonisi}},
  \bibinfo{editor}{\bibfnamefont{J.~R.} \bibnamefont{Schrieffer}},
  \bibnamefont{and} \bibinfo{editor}{\bibfnamefont{M.~L.}
  \bibnamefont{Chiofalo}} (\bibinfo{publisher}{IOS},
  \bibinfo{address}{Amsterdam}, \bibinfo{year}{1999}), Proceedings of the
  CXXXVI International School of Physics ``E. Fermi'', Varenna (Italy), 1997.

\bibitem[{\citenamefont{Abrikosov et~al.}(1993)\citenamefont{Abrikosov,
  Campuzano, and Gofron}}]{Abrikosov:93}
\bibinfo{author}{\bibfnamefont{A.~A.} \bibnamefont{Abrikosov}},
  \bibinfo{author}{\bibfnamefont{J.}~\bibnamefont{Campuzano}},
  \bibnamefont{and} \bibinfo{author}{\bibfnamefont{K.}~\bibnamefont{Gofron}},
  \bibinfo{journal}{Physica C} \textbf{\bibinfo{volume}{214}},
  \bibinfo{pages}{73} (\bibinfo{year}{1993}).

\bibitem[{\citenamefont{Dessau et~al.}(1993)\citenamefont{Dessau, Shen, King,
  Marshall, Lombardo, Dickinson, Loeser, {DiCarlo}, Park, Kapitulnik
  et~al.}}]{Dessau:93}
\bibinfo{author}{\bibfnamefont{D.~S.} \bibnamefont{Dessau}},
  \bibinfo{author}{\bibfnamefont{Z.}~\bibnamefont{Shen}},
  \bibinfo{author}{\bibfnamefont{D.~M.} \bibnamefont{King}},
  \bibinfo{author}{\bibfnamefont{D.~S.} \bibnamefont{Marshall}},
  \bibinfo{author}{\bibfnamefont{L.~W.} \bibnamefont{Lombardo}},
  \bibinfo{author}{\bibfnamefont{P.~H.} \bibnamefont{Dickinson}},
  \bibinfo{author}{\bibfnamefont{A.~G.} \bibnamefont{Loeser}},
  \bibinfo{author}{\bibfnamefont{J.}~\bibnamefont{{DiCarlo}}},
  \bibinfo{author}{\bibfnamefont{C.}~\bibnamefont{Park}},
  \bibinfo{author}{\bibfnamefont{A.}~\bibnamefont{Kapitulnik}},
  \bibnamefont{et~al.}, \bibinfo{journal}{Phys. Rev. Lett.}
  \textbf{\bibinfo{volume}{71}}, \bibinfo{pages}{2781} (\bibinfo{year}{1993}).

\bibitem[{\citenamefont{Gofron et~al.}(1994)\citenamefont{Gofron, Campuzano,
  Abrikosov, Lindroos, Bansil, Ding, Koelling, and Dabrowski}}]{Gofron:94}
\bibinfo{author}{\bibfnamefont{K.}~\bibnamefont{Gofron}},
  \bibinfo{author}{\bibfnamefont{J.~C.} \bibnamefont{Campuzano}},
  \bibinfo{author}{\bibfnamefont{A.~A.} \bibnamefont{Abrikosov}},
  \bibinfo{author}{\bibfnamefont{M.}~\bibnamefont{Lindroos}},
  \bibinfo{author}{\bibfnamefont{A.}~\bibnamefont{Bansil}},
  \bibinfo{author}{\bibfnamefont{H.}~\bibnamefont{Ding}},
  \bibinfo{author}{\bibfnamefont{D.}~\bibnamefont{Koelling}}, \bibnamefont{and}
  \bibinfo{author}{\bibfnamefont{B.}~\bibnamefont{Dabrowski}},
  \bibinfo{journal}{Phys. Rev. Lett.} \textbf{\bibinfo{volume}{73}},
  \bibinfo{pages}{3302} (\bibinfo{year}{1994}).

\bibitem[{\citenamefont{Hodges et~al.}(1971)\citenamefont{Hodges, Smith, and
  Wilkins}}]{Hodges:71}
\bibinfo{author}{\bibfnamefont{C.}~\bibnamefont{Hodges}},
  \bibinfo{author}{\bibfnamefont{H.}~\bibnamefont{Smith}}, \bibnamefont{and}
  \bibinfo{author}{\bibfnamefont{J.~W.} \bibnamefont{Wilkins}},
  \bibinfo{journal}{Phys. Rev. B} \textbf{\bibinfo{volume}{4}},
  \bibinfo{pages}{302} (\bibinfo{year}{1971}).

\bibitem[{\citenamefont{Millis et~al.}(1990)\citenamefont{Millis, Monien, and
  Pines}}]{Millis:90}
\bibinfo{author}{\bibfnamefont{A.~J.} \bibnamefont{Millis}},
  \bibinfo{author}{\bibfnamefont{H.}~\bibnamefont{Monien}}, \bibnamefont{and}
  \bibinfo{author}{\bibfnamefont{D.}~\bibnamefont{Pines}},
  \bibinfo{journal}{Phys. Rev. B} \textbf{\bibinfo{volume}{42}},
  \bibinfo{pages}{167} (\bibinfo{year}{1990}).

\bibitem[{\citenamefont{Emery and Kivelson}(1995)}]{Emery:95a}
\bibinfo{author}{\bibfnamefont{V.~J.} \bibnamefont{Emery}} \bibnamefont{and}
  \bibinfo{author}{\bibfnamefont{S.~A.} \bibnamefont{Kivelson}},
  \bibinfo{journal}{Nature} \textbf{\bibinfo{volume}{374}},
  \bibinfo{pages}{434} (\bibinfo{year}{1995}).

\bibitem[{\citenamefont{Castellani et~al.}(1995)\citenamefont{Castellani, {Di
  Castro}, and Grilli}}]{Castellani:95}
\bibinfo{author}{\bibfnamefont{C.}~\bibnamefont{Castellani}},
  \bibinfo{author}{\bibfnamefont{C.}~\bibnamefont{{Di Castro}}},
  \bibnamefont{and} \bibinfo{author}{\bibfnamefont{M.}~\bibnamefont{Grilli}},
  \bibinfo{journal}{Phys. Rev. Lett.} \textbf{\bibinfo{volume}{75}},
  \bibinfo{pages}{4650} (\bibinfo{year}{1995}).

\bibitem[{\citenamefont{Emery and Kivelson}(1993)}]{Emery:93}
\bibinfo{author}{\bibfnamefont{V.~J.} \bibnamefont{Emery}} \bibnamefont{and}
  \bibinfo{author}{\bibfnamefont{S.~A.} \bibnamefont{Kivelson}},
  \bibinfo{journal}{Physica C} \textbf{\bibinfo{volume}{209}},
  \bibinfo{pages}{597} (\bibinfo{year}{1993}).

\bibitem[{\citenamefont{Chakravarty et~al.}(1993)\citenamefont{Chakravarty,
  Sudb{\o}, Anderson, and Strong}}]{Chakravarty:93}
\bibinfo{author}{\bibfnamefont{S.}~\bibnamefont{Chakravarty}},
  \bibinfo{author}{\bibfnamefont{A.}~\bibnamefont{Sudb{\o}}},
  \bibinfo{author}{\bibfnamefont{P.~W.} \bibnamefont{Anderson}},
  \bibnamefont{and} \bibinfo{author}{\bibfnamefont{S.}~\bibnamefont{Strong}},
  \bibinfo{journal}{Science} \textbf{\bibinfo{volume}{261}},
  \bibinfo{pages}{337} (\bibinfo{year}{1993}).

\bibitem[{\citenamefont{Angilella et~al.}(1999)\citenamefont{Angilella, Pucci,
  Siringo, and Sudb{\o}}}]{Angilella:99}
\bibinfo{author}{\bibfnamefont{G.~G.~N.} \bibnamefont{Angilella}},
  \bibinfo{author}{\bibfnamefont{R.}~\bibnamefont{Pucci}},
  \bibinfo{author}{\bibfnamefont{F.}~\bibnamefont{Siringo}}, \bibnamefont{and}
  \bibinfo{author}{\bibfnamefont{A.}~\bibnamefont{Sudb{\o}}},
  \bibinfo{journal}{Phys. Rev. B} \textbf{\bibinfo{volume}{59}},
  \bibinfo{pages}{1339} (\bibinfo{year}{1999}).

\bibitem[{\citenamefont{Gonz{\'a}lez et~al.}(1996)\citenamefont{Gonz{\'a}lez,
  Guinea, and {M. A. H. Vozmediano}}}]{Gonzalez:96}
\bibinfo{author}{\bibfnamefont{J.}~\bibnamefont{Gonz{\'a}lez}},
  \bibinfo{author}{\bibfnamefont{F.}~\bibnamefont{Guinea}}, \bibnamefont{and}
  \bibinfo{author}{\bibnamefont{{M. A. H. Vozmediano}}},
  \bibinfo{journal}{Europhys. Lett.} \textbf{\bibinfo{volume}{34}},
  \bibinfo{pages}{711} (\bibinfo{year}{1996}).

\bibitem[{\citenamefont{{Gonz\'alez}}(2000)}]{Gonzalez:01}
\bibinfo{author}{\bibfnamefont{J.}~\bibnamefont{{Gonz\'alez}}},
  \bibinfo{journal}{Phys. Rev. B} \textbf{\bibinfo{volume}{63}},
  \bibinfo{pages}{45114} (\bibinfo{year}{2000}).

\bibitem[{\citenamefont{Irkhin et~al.}(2001{\natexlab{b}})\citenamefont{Irkhin,
  Katanin, and Katsnelson}}]{Irkhin:01b}
\bibinfo{author}{\bibfnamefont{V.}~\bibnamefont{Irkhin}},
  \bibinfo{author}{\bibfnamefont{A.~A.} \bibnamefont{Katanin}},
  \bibnamefont{and} \bibinfo{author}{\bibfnamefont{M.~I.}
  \bibnamefont{Katsnelson}}, \bibinfo{journal}{...}
  \textbf{\bibinfo{volume}{...}}, \bibinfo{pages}{...}
  (\bibinfo{year}{2001}{\natexlab{b}}), \bibinfo{note}{preprint {\tt
  cond-mat/0110516}}.

\bibitem[{Mor()}]{Morse}
\bibinfo{note}{Since the effective mass tensor is non-degenerate on every
  critical point of $\varepsilon_\bk$, i.e. $\det\partial_{k_i k_j}
  \varepsilon_\bk \neq0$ when $\partial_{k_i} \varepsilon_\bk =0$,
  $\varepsilon_\bk$ plays the role of a Morse function over the 1BZ [see, e.g.,
  J. Milnor, \emph{Morse theory,} Ann. Math. Studies 51 (Princeton University
  Press, Princeton, 1963)].}

\bibitem[{\citenamefont{Xing et~al.}(1991)\citenamefont{Xing, Liu, and
  Gong}}]{Xing:91}
\bibinfo{author}{\bibfnamefont{D.~Y.} \bibnamefont{Xing}},
  \bibinfo{author}{\bibfnamefont{M.}~\bibnamefont{Liu}}, \bibnamefont{and}
  \bibinfo{author}{\bibfnamefont{C.~D.} \bibnamefont{Gong}},
  \bibinfo{journal}{Phys. Rev. B} \textbf{\bibinfo{volume}{44}},
  \bibinfo{pages}{12525} (\bibinfo{year}{1991}).

\bibitem[{\citenamefont{Gradshteyn and Ryzhik}(1994)}]{GR}
\bibinfo{author}{\bibfnamefont{I.~S.} \bibnamefont{Gradshteyn}}
  \bibnamefont{and} \bibinfo{author}{\bibfnamefont{I.~M.}
  \bibnamefont{Ryzhik}}, \emph{\bibinfo{title}{Table of Integrals, Series, and
  Products}} (\bibinfo{publisher}{Academic Press}, \bibinfo{address}{Boston},
  \bibinfo{year}{1994}), \bibinfo{edition}{5th} ed.

\bibitem[{\citenamefont{Abramowitz and Stegun}(1964)}]{AS}
\bibinfo{editor}{\bibfnamefont{M.}~\bibnamefont{Abramowitz}} \bibnamefont{and}
  \bibinfo{editor}{\bibfnamefont{I.~A.} \bibnamefont{Stegun}}, eds.,
  \emph{\bibinfo{title}{Handbook of Mathematical Functions}}
  (\bibinfo{publisher}{Dover}, \bibinfo{address}{New York},
  \bibinfo{year}{1964}).

\bibitem[{\citenamefont{Chubukov and Morr}(1997)}]{Chubukov:97a}
\bibinfo{author}{\bibfnamefont{A.~V.} \bibnamefont{Chubukov}} \bibnamefont{and}
  \bibinfo{author}{\bibfnamefont{D.~K.} \bibnamefont{Morr}},
  \bibinfo{journal}{Phys. Rep.} \textbf{\bibinfo{volume}{288}},
  \bibinfo{pages}{355} (\bibinfo{year}{1997}).

\bibitem[{\citenamefont{Abrikosov et~al.}(1975)\citenamefont{Abrikosov, Gorkov,
  and Dzyaloshinski}}]{AGD}
\bibinfo{author}{\bibfnamefont{A.~A.} \bibnamefont{Abrikosov}},
  \bibinfo{author}{\bibfnamefont{L.~P.} \bibnamefont{Gorkov}},
  \bibnamefont{and} \bibinfo{author}{\bibfnamefont{I.~E.}
  \bibnamefont{Dzyaloshinski}}, \emph{\bibinfo{title}{Methods of Quantum Field
  Theory in Statistical Physics}} (\bibinfo{publisher}{Dover},
  \bibinfo{address}{New York}, \bibinfo{year}{1975}).

\bibitem[{\citenamefont{Loeser et~al.}(1996)\citenamefont{Loeser, Dessau, and
  Shen}}]{Loeser:96}
\bibinfo{author}{\bibfnamefont{A.~G.} \bibnamefont{Loeser}},
  \bibinfo{author}{\bibfnamefont{D.~S.} \bibnamefont{Dessau}},
  \bibnamefont{and} \bibinfo{author}{\bibfnamefont{Z.}~\bibnamefont{Shen}},
  \bibinfo{journal}{Physica C} \textbf{\bibinfo{volume}{263}},
  \bibinfo{pages}{208} (\bibinfo{year}{1996}).

\bibitem[{\citenamefont{Landau and Lifshitz}(1981)}]{Landau:X}
\bibinfo{author}{\bibfnamefont{L.~D.} \bibnamefont{Landau}} \bibnamefont{and}
  \bibinfo{author}{\bibfnamefont{E.~M.} \bibnamefont{Lifshitz}},
  \emph{\bibinfo{title}{Physical Kinetics}}, vol.~\bibinfo{volume}{10} of
  \emph{\bibinfo{series}{Course of Theoretical Physics}}
  (\bibinfo{publisher}{Pergamon}, \bibinfo{address}{New York},
  \bibinfo{year}{1981}).

\bibitem[{\citenamefont{Hlubina and Rice}(1995)}]{Hlubina:95}
\bibinfo{author}{\bibfnamefont{R.}~\bibnamefont{Hlubina}} \bibnamefont{and}
  \bibinfo{author}{\bibfnamefont{T.~M.} \bibnamefont{Rice}},
  \bibinfo{journal}{Phys. Rev. B} \textbf{\bibinfo{volume}{51}},
  \bibinfo{pages}{9253} (\bibinfo{year}{1995}).

\bibitem[{\citenamefont{Varlamov and Pantsulaya}(1985)}]{Varlamov:85}
\bibinfo{author}{\bibfnamefont{A.~A.} \bibnamefont{Varlamov}} \bibnamefont{and}
  \bibinfo{author}{\bibfnamefont{A.~V.} \bibnamefont{Pantsulaya}},
  \bibinfo{journal}{Sov. Phys. JETP} \textbf{\bibinfo{volume}{62}},
  \bibinfo{pages}{1263} (\bibinfo{year}{1985}), \bibinfo{note}{[Zh. Eksp. Teor.
  Fiz. {\bf 89}, 2188 (1985)]}.

\bibitem[{\citenamefont{Sun and Maki}(1995)}]{Sun:95}
\bibinfo{author}{\bibfnamefont{Y.}~\bibnamefont{Sun}} \bibnamefont{and}
  \bibinfo{author}{\bibfnamefont{K.}~\bibnamefont{Maki}},
  \bibinfo{journal}{Phys. Rev. B} \textbf{\bibinfo{volume}{51}},
  \bibinfo{pages}{6059} (\bibinfo{year}{1995}).

\bibitem[{\citenamefont{Lee}(1997)}]{Lee:97}
\bibinfo{author}{\bibfnamefont{P.~A.} \bibnamefont{Lee}},
  \bibinfo{journal}{Science} \textbf{\bibinfo{volume}{277}},
  \bibinfo{pages}{50} (\bibinfo{year}{1997}).

\bibitem[{\citenamefont{Lee and Wen}(1997)}]{Lee:97a}
\bibinfo{author}{\bibfnamefont{P.~A.} \bibnamefont{Lee}} \bibnamefont{and}
  \bibinfo{author}{\bibfnamefont{X.~G.} \bibnamefont{Wen}},
  \bibinfo{journal}{Phys. Rev. Lett.} \textbf{\bibinfo{volume}{78}},
  \bibinfo{pages}{4111} (\bibinfo{year}{1997}).

\bibitem[{\citenamefont{Balents et~al.}(1998)\citenamefont{Balents, Fisher, and
  Nayak}}]{Balents:98}
\bibinfo{author}{\bibfnamefont{L.}~\bibnamefont{Balents}},
  \bibinfo{author}{\bibfnamefont{M.~P.~A.} \bibnamefont{Fisher}},
  \bibnamefont{and} \bibinfo{author}{\bibfnamefont{C.}~\bibnamefont{Nayak}},
  \bibinfo{journal}{Int. J. Mod. Phys. B} \textbf{\bibinfo{volume}{12}},
  \bibinfo{pages}{1033} (\bibinfo{year}{1998}).

\bibitem[{\citenamefont{Khveshchenko et~al.}(2001)\citenamefont{Khveshchenko,
  Yashenkin, and Gornyi}}]{Khveshchenko:01}
\bibinfo{author}{\bibfnamefont{D.~V.} \bibnamefont{Khveshchenko}},
  \bibinfo{author}{\bibfnamefont{A.~G.} \bibnamefont{Yashenkin}},
  \bibnamefont{and} \bibinfo{author}{\bibfnamefont{I.~V.}
  \bibnamefont{Gornyi}}, \bibinfo{journal}{Phys. Rev. Lett.}
  \textbf{\bibinfo{volume}{86}}, \bibinfo{pages}{4668} (\bibinfo{year}{2001}).

\bibitem[{\citenamefont{Yashenkin et~al.}(2001)\citenamefont{Yashenkin,
  Atkinson, Gornyi, Hirschfeld, and Khveshchenko}}]{Yashenkin:01}
\bibinfo{author}{\bibfnamefont{A.~G.} \bibnamefont{Yashenkin}},
  \bibinfo{author}{\bibfnamefont{W.~A.} \bibnamefont{Atkinson}},
  \bibinfo{author}{\bibfnamefont{I.~V.} \bibnamefont{Gornyi}},
  \bibinfo{author}{\bibfnamefont{P.~J.} \bibnamefont{Hirschfeld}},
  \bibnamefont{and} \bibinfo{author}{\bibfnamefont{D.~V.}
  \bibnamefont{Khveshchenko}}, \bibinfo{journal}{Phys. Rev. Lett.}
  \textbf{\bibinfo{volume}{86}}, \bibinfo{pages}{5982} (\bibinfo{year}{2001}).

\bibitem[{\citenamefont{Cappelluti and
  Pietronero}(1996{\natexlab{a}})}]{Cappelluti:96a}
\bibinfo{author}{\bibfnamefont{E.}~\bibnamefont{Cappelluti}} \bibnamefont{and}
  \bibinfo{author}{\bibfnamefont{L.}~\bibnamefont{Pietronero}},
  \bibinfo{journal}{Phys. Rev. B} \textbf{\bibinfo{volume}{53}},
  \bibinfo{pages}{932} (\bibinfo{year}{1996}{\natexlab{a}}).

\bibitem[{\citenamefont{Cappelluti and
  Pietronero}(1996{\natexlab{b}})}]{Cappelluti:96b}
\bibinfo{author}{\bibfnamefont{E.}~\bibnamefont{Cappelluti}} \bibnamefont{and}
  \bibinfo{author}{\bibfnamefont{L.}~\bibnamefont{Pietronero}},
  \bibinfo{journal}{Europhys. Lett.} \textbf{\bibinfo{volume}{36}},
  \bibinfo{pages}{619} (\bibinfo{year}{1996}{\natexlab{b}}).

\bibitem[{\citenamefont{Liu et~al.}(1997)\citenamefont{Liu, Xing, and
  Wang}}]{Liu:97}
\bibinfo{author}{\bibfnamefont{M.}~\bibnamefont{Liu}},
  \bibinfo{author}{\bibfnamefont{D.~Y.} \bibnamefont{Xing}}, \bibnamefont{and}
  \bibinfo{author}{\bibfnamefont{Z.~D.} \bibnamefont{Wang}},
  \bibinfo{journal}{Phys. Rev. B} \textbf{\bibinfo{volume}{55}},
  \bibinfo{pages}{3181} (\bibinfo{year}{1997}).

\bibitem[{\citenamefont{Hancock}(1910)}]{Hancock:10}
\bibinfo{author}{\bibfnamefont{H.}~\bibnamefont{Hancock}},
  \emph{\bibinfo{title}{Lectures on the Theory of Elliptic Functions}},
  vol.~\bibinfo{volume}{1} (\bibinfo{publisher}{J. Wiley \& Sons},
  \bibinfo{address}{New York}, \bibinfo{year}{1910}).

\end{thebibliography}

\newpage

\newpage

\newpage

\newpage

\end{document}